\documentclass[aps,pra,twocolumn,superscriptaddress,showpacs,preprintnumbers,amsmath,amssymb]{revtex4-2}

\usepackage{bm}
\usepackage{graphicx}
\usepackage{dcolumn}
\usepackage{color} 
\usepackage{bbm}
\usepackage[normalem]{ulem}
\usepackage{isotope}
\usepackage{amssymb}

\usepackage[colorlinks=true, linkcolor=blue, citecolor=blue, urlcolor=blue]{hyperref}
\newcommand{\ii}{\mathrm{i}}
\newcommand{\ee}{\mathrm{e}}
\renewcommand{\vec}[1]{\mbox{\boldmath $#1$}} 

\begin{document}

\title{Floquet-engineered nondispersive wave packets in helium under combined periodic and static fields} 

\author{Alejandro Gonz{\'a}lez-Melan}
\email[]{alejandro.gonzalez.melan@correounivalle.edu.co}
\affiliation{Physics Department, Universidad del Valle, Cali, Colombia}

\author{John Fitzgerald Le{\'o}n Ocampo}
\email[]{jhon john.fitzgerald.leon@correounivalle.edu.co}
\affiliation{Physics Department, Universidad del Valle, Cali, Colombia}

\author{Javier Madro\~nero}
\email[]{javier.madronero@correounivalle.edu.co}
\affiliation{Physics Department, Universidad del Valle, Cali, Colombia\\
Centre for Bioinformatics and Photonics (CiBioFi), Universidad del 
Valle, Cali, Colombia}

\date{\today}

\begin{abstract}
We report the numerical observation of two-electron nondispersive wave packets in fully three-dimensional helium subjected to a linearly polarized monochromatic field.
These localized quantum states follow periodic trajectories of the classical three-body Coulomb system without dispersion.
The system is modeled using a spectral configuration interaction approach based on Coulomb-Sturmian functions, combined with Floquet theory and complex scaling.
The resulting quasienergy spectrum reveals the existence of long-lived Floquet states arising from near-resonant coupling between frozen planet configurations.
\end{abstract}

\pacs{32.80.Qk, 31.15.ac, 32.80.Rm, 05.45.Mt}
\maketitle

\section{Introduction}
\label{sec:intro}

Understanding and controlling the dynamics of correlated electrons in atoms and molecules remains a central challenge in modern atomic, molecular, and optical physics, 
where strong-field and ultrafast phenomena are governed by multielectron interactions \cite{Ossiander2017,Busto2018}. 
Electron-electron correlation plays a fundamental role in a variety of non-equilibrium processes, including ionization, autoionization, and field-induced excitation, 
and lies at the core of attosecond electronic motion in driven systems \cite{Beaulieu2017}.

Recent advances in the control of optical field transients have enabled coherent generation of isolated attosecond pulses \cite{Yang2021}, while attosecond spectroscopy has emerged as a powerful tool to investigate 
electronic and nuclear dynamics with unprecedented temporal resolution in atoms, molecules, and solids \cite{BorregoVarillas2022}. Furthermore, coherent population transfer in helium has been observed via Rabi oscillations 
driven by extreme ultraviolet pulses \cite{Nandi2022}, demonstrating the applicability of attosecond science in the study of strong-field processes in two-electron atoms.

In the broader context of few-electron systems, the helium atom stands out as prototypical case for investigating correlated quantum dynamics. From a classical perspective, helium is a non-integrable three-body Coulomb system 
with a rich phase space structure exhibiting both chaotic and regular motions \cite{Richter1993,Tanner2000}. This lack of integrability manifests in its quantum spectrum through the appearance of highly correlated doubly 
excited states, quantum chaos signatures, and complex resonance structures \cite{Madden1963,Byun2007,Ericson1960,Ericson1963}. 
The influence of electronic correlation has also been observed in measurable quantities such as photon momentum transfer in single ionization, where mean-field models fail to capture the energy dependence seen in experiments \cite{Fang2025}. Understanding these correlated dynamics is crucial not only for fundamental science but also for potential applications in quantum computing \cite{Blatt2008, Georgescu2014} and in the simulation and design of correlated materials \cite{Ma2020}.

A particularly interesting class of doubly excited states is associated with the so-called \textit{frozen planet configuration} (FPC), a highly asymmetric classical configuration in which both electrons remain on the 
same side of the nucleus \cite{Richter1990,Richter1992}. The quantum counterparts of these configurations are long-lived doubly excited states known as \textit{frozen planet states} (FPS). 
Under near-resonant periodic driving, such states can transform into \textit{nondispersive wave packets} (NDWP), that is, localized quantum states which follow periodic classical trajectories without 
spreading \cite{Schlagheck1999,Buchleitner2002}.

NDWP have been identified in reduced-dimensionality models of helium, including one-dimensional \cite{Schlagheck1999} and planar systems \cite{Madronero2008,GonzalezMelan2020}. 
However, their existence in fully three-dimensional helium remains largely unexplored because of the substantial numerical complexity involved, which is a challenge that we address in the present work. 
The recent development of attosecond four-wave-mixing spectroscopy has enabled the extraction of lifetimes of doubly excited states directly in the time domain, reinforcing the importance of a detailed theoretical 
understanding of their structure and decay mechanisms \cite{Rupprecht2024}.  This capability offers new possibilities for experimental validation of theoretical predictions and provides deeper insights into the dynamics 
of highly correlated systems.

In this work, we perform a full-dimensional theoretical study of helium driven by a linearly polarized monochromatic field.
The dynamics is investigated using a spectral method, where the time-dependent Schrödinger equation (TDSE) is solved in the basis of eigenstates of the unperturbed Hamiltonian,
combined with Floquet theory \cite{Floquet1,shirley1}. This representation leads to a significant reduction in the size of the Floquet matrices for the driven system.
The field-free Hamiltonian is diagonalized in a configuration interaction (CI) basis \cite{lagmago:phd99,eiglsperger:3d,eiglspergerdiss}, where the radial one-electron functions are Coulomb-Sturmian functions with different nonlinear parameters for each electron, chosen to efficiently target specific energy regimes.
This enables an efficient representation of asymmetrically excited configurations, such as the frozen planet states, using a comparatively small basis.

Additionally, it is of interest to examine whether the stability of such wave packets can be enhanced by external control mechanisms. Motivated by classical predictions that suggest a stabilizing 
effect of a weak static electric field on FPC trajectories \cite{Schlagheck1998}, we analyze the effect of a static field component added along the polarization axis. Our results show that this additional 
field can increase the lifetime of selected Floquet states. 
This provides a viable route for coherence enhancement in strongly correlated wave packet dynamics and complements previous studies based on adiabatic approaches to frozen-planet resonances \cite{Grozdanov2020}, by extending such stabilization strategies to a fully quantum, three-dimensional framework.

This paper is organized as follows. In Section~\ref{sec:theory}, we present the theoretical framework, including the field-free helium atom, the Floquet formalism for periodic driving, 
the complex scaling method and the classical FPC. Section~\ref{sec:results} contains the numerical results: we begin by identifying the frozen planet states of the unperturbed atom, 
then examine their behavior under periodic driving. The influence of a static electric field is analyzed both in isolation and in combination with the periodic drive, and we conclude 
with a discussion of prospects for experimental detection of nondispersive wave packets. Finally, Section~\ref{sec:conclusions} summarizes our findings.

\section{Theory}
\label{sec:theory}

In the center-of-mass frame and within the infinite nuclear mass approximation, 
the dynamics of a two-electron atom subjected to a linearly polarized electromagnetic field is governed by the Hamiltonian (atomic units are used throughout)
\begin{eqnarray}\label{ec:Ham_field}
H(t) &=& H_0 + (F \cos \omega t + F_{\textrm{st}})(x_1 + x_2)\,,
\end{eqnarray}
where $F$ and $\omega$ denote the amplitude and frequency of the monochromatic driving field, and $F_{\textrm{st}}$ is the strength of the additional static electric field. Both fields are taken to 
be spatially uniform and oriented along the $x$-axis. The coupling is introduced in the length gauge under the dipole approximation.

The field-free Hamiltonian $H_0$ reads
\begin{equation}\label{ec:Ham_free}
H_0 = \frac{\vec{p}_1^{\;2}}{2} + \frac{\vec{p}_2^{\;2}}{2} - \frac{Z}{r_1} - 
\frac{Z}{r_2} + \frac{\gamma}{r_{12}}\,,
\end{equation}
where $Z$ is the nuclear charge and $\gamma$ characterizes the electron-electron interaction. 
For helium, we set $\gamma = 1$ and $Z = 2$. The vectors $\vec{r}_1$ and $\vec{r}_2$ denote the positions of the electrons relative to the nucleus, and $\vec{p}_1$, $\vec{p}_2$ are their corresponding conjugate momenta.

The TDSE associated with the Hamiltonian~(\ref{ec:Ham_field}) is solved using a spectral method based on Floquet theory and configuration interaction techniques~\cite{GonzalezMelan2020}. 
The field-free Hamiltonian $H_0$ is first represented in a configuration interaction basis constructed from products of Coulomb-Sturmian functions and diagonalized for fixed total angular momentum $L$, yielding a discrete set of unperturbed eigenstates. 
The wave function is then expanded in this \emph{atomic basis} to represent the evolution of the system under various field configurations, including periodic driving, a static electric field, or both combined.

The remainder of this section provides a detailed description of the theoretical framework used in our study. 
We begin with the field-free configuration interaction approach for helium based on Coulomb-Sturmian functions, then introduce the Floquet formalism for time-periodic fields, followed by the complex scaling method for extracting resonances.
we conclude with an outline of the classical FPC, which underpins the correlated dynamics explored in the quantum regime.

\subsection{Field-free helium atom}
\label{subsec:field-free-helium}

The wave function $|\varphi^L_j\rangle = |\varphi^L_j(\vec{r}_1,\vec{r}_2)\rangle$ of the helium atom with total angular momentum $L$ and energy $\epsilon^L_j$ satisfies the stationary 
Schrödinger equation
\begin{equation}\label{ec:TISE}
H_0|\varphi^L_j\rangle = \epsilon^L_j|\varphi^L_j\rangle\,,
\end{equation}
and is expanded in a configuration interaction basis \cite{foumouo06,eiglsperger:3d} as follows:
\begin{eqnarray}\label{ec:Expansion}
\varphi^L_j(\vec{r}_1,\vec{r}_2) &=& \sum_{\epsilon_{12},\pi} {\sum_{M,l_1,l_2}}^\pi \sum_s \sum_{n_1,n_2} 
\varphi^{l_1,l_2,L,M,\epsilon_{12}}_{\kappa_{1s},\kappa_{2s},n_1,n_2,j} \,
\beta^{l_1,l_2}_{n_1,n_2} \nonumber \\
&\times& \mathcal{A} \,
F^{l_1,l_2,L,M}_{\kappa_{1s},\kappa_{2s},n_1,n_2}(\vec{r}_1,\vec{r}_2)\,,
\end{eqnarray}
where
\begin{eqnarray}
F^{l_1,l_2,L,M}_{\kappa_{1s},\kappa_{2s},n_1,n_2}(\vec{r}_1,\vec{r}_2) &=& 
\frac{S^{(\kappa_{1s})}_{n_1,l_1}(r_1)}{r_1}
\frac{S^{(\kappa_{2s})}_{n_2,l_2}(r_2)}{r_2} \nonumber \\
&\times&\Lambda^{L,M}_{l_1,l_2}(\hat{r}_1, \hat{r}_2)\,,
\end{eqnarray}
with $\hat{r} \equiv (\theta,\phi)$ denoting the angular coordinates and $M$ the projection of the total orbital angular momentum along the $z$-axis.

The factor
\begin{equation}
\beta^{l_1,l_2}_{n_1,n_2} = 1 + \left(\frac{1}{\sqrt{2}} - 1\right)\delta_{n_1,n_2}\,\delta_{l_1,l_2}\,,
\end{equation}
eliminates redundant basis functions arising from identical one-electron states. The operator
\begin{equation}
\mathcal{A} = \frac{1 + (-1)^{l_1 + l_2 - L}\epsilon_{12} P}{\sqrt{2}}\,,
\end{equation}
projects onto singlet ($\epsilon_{12} = +1$) or triplet ($\epsilon_{12} = -1$) spin subspaces, where $P$ exchanges simultaneously $(\kappa_{1s},n_1,l_1)$ with $(\kappa_{2s},n_2,l_2)$. 
The symbol ${\sum}^\pi$ in (\ref{ec:Expansion}) indicates that the sum runs over quantum numbers compatible with the parity $\pi$, distinguishing between natural parity states ($\pi = (-1)^L$) and 
unnatural parity states ($\pi = (-1)^{L+1}$).

The radial part of the wave function is expanded in one-electron Coulomb-Sturmian functions $S^{(\kappa)}_{n,l}(r)$ \cite{rotenberg70,huens97}
\begin{equation}
S^{(\kappa)}_{n,l}(r) = N^{\kappa}_{n,l}\, r^{l+1} e^{-\kappa r} L^{(2l+1)}_{n-l-1}(2\kappa r)\,,
\end{equation}
which satisfy the orthogonality relation
\begin{equation}
\int_0^\infty dr\, S^{(\kappa)}_{n,l}(r)\frac{1}{r}S^{(\kappa)}_{n',l}(r) = \frac{\kappa}{n}\delta_{n,n'}\,.
\end{equation}
Here, $l$ is the orbital angular momentum, $n$ the radial quantum number, $\kappa$ a nonlinear dilation parameter, $L^{(\alpha)}_m(x)$ an associated Laguerre polynomial, and
\begin{equation}
N^{\kappa}_{n,l} = \sqrt{\frac{\kappa(n - l - 1)!}{n(n + l)!}}\,,
\end{equation}
is the normalization constant.

The angular part of the wave function is expanded in terms of bipolar spherical harmonics \cite{Varschalovich08}:
\begin{eqnarray}
\Lambda^{L,M}_{l_1,l_2}(\hat{r}_1,\hat{r}_2)=\sum_{m_1,m_2}(-1)^{l_1-l_2+M}\sqrt{2L+1}\nonumber\\
\times\left(\begin{array}{ccc}
l_1 & l_2 & L\\
m_1 & m_2 & -M
\end{array}\right)
 Y_{l_1,m_1}(\hat{r}_1)Y_{l_2,m_2}(\hat{r}_2)\,,
\end{eqnarray}
with the orthonormalization relation
\begin{eqnarray}
\int d\hat{r}_1d\hat{r}_2\, {\Lambda^{L,M}_{l_1,l_2}}^*(\hat{r}_1,\hat{r}_2) \Lambda^{L',M'}_{l'_1,l'_2}(\hat{r}_1,\hat{r}_2)\nonumber\\
=\delta_{l_1,l'_1}\,\delta_{l_2,l'_2}\,\delta_{L,L'}\,\delta_{M,M'}\,,
\end{eqnarray}
where $Y_{l,m}(\hat{r})$ are the spherical harmonics and the symbol in parentheses is the Wigner 3$j$-symbol \cite{Varschalovich08}.

Since $r_{12}$ is not an explicit coordinate in the CI basis, the electron-electron interaction is treated via the multipole expansion:
\begin{eqnarray}\label{ec:mult-ee}
\frac{1}{r_{12}} = \sum_{q=0}^\infty \sum_{p=-q}^q \frac{4\pi}{2q+1} \frac{r^q_<}{r^{q+1}_>} Y^*_{q,p}(\hat{r}_1)Y_{q,p}(\hat{r}_2)\,,
\end{eqnarray}
where $r_< = \min(r_1,r_2)$ and $r_> = \max(r_1,r_2)$.

The function $S^{(\kappa)}_{n,l}(r)$ coincides with the hydrogenic radial eigenfunction when $\kappa = Z/n$. The parameter $\kappa$ can therefore be optimized to improve convergence for different spectral regions. 
For large values of $\kappa$, the function is spatially localized near the origin, while for small $\kappa$, it extends over large distances. 
Thus, for describing symmetrically excited states, all radial functions can be constructed with equal $\kappa$ values. However, in the case of asymmetrically excited states—such as 
the frozen planet states discussed in Section~\ref{subsec:fps-identification}—an efficient description is obtained by assigning different dilation parameters to each electron. The inner electron, localized near the nucleus, 
is described with a large $\kappa$, whereas the outer electron is represented with small $\kappa$ functions. This asymmetric construction allows us to represent relevant configurations with a reduced 
number of basis functions, reducing the overall size of the CI expansion as compared to standard approaches.

For each angular momentum pair $(l_1,l_2)$, we define one or several basis sets characterized by the parameters $\{\kappa_{1s}, N_{1s}^{\textrm{min}}, N_{1s}^{\textrm{max}}, \kappa_{2s}, N_{2s}^{\textrm{min}}, N_{2s}^{\textrm{max}}\}$. 
The radial quantum numbers $n_1$ and $n_2$ satisfy
\[
l_1 + N_{1s}^{\textrm{min}} \leq n_1 \leq l_1 + N_{1s}^{\textrm{max}}, \quad 
l_2 + N_{2s}^{\textrm{min}} \leq n_2 \leq l_2 + N_{2s}^{\textrm{max}}.
\]
All redundant combinations of Coulomb-Sturmian functions resulting from particle exchange are excluded from the expansion~(\ref{ec:Expansion}) by applying a single value decomposition (SVD) to the basis \cite{eiglsperger:3d}.

\subsection{Helium atom under periodic driving}
\label{subsec:periodic-driving}

Within the Floquet formalism \cite{Floquet1,shirley1}, the solutions $|\psi_j(t)\rangle = |\psi_j(\vec{r}_1,\vec{r}_2,t)\rangle$ of the TDSE associated with the Hamiltonian~(\ref{ec:Ham_field}) 
can be expressed in terms of time-periodic functions as
\begin{eqnarray}\nonumber
|\psi_j(t)\rangle = \sum_\alpha 
C_{j,\alpha}\,\ee^{-i\varepsilon_\alpha t}|\phi_\alpha(t)\rangle,\quad |\phi_\alpha(t+T)\rangle = |\phi_\alpha(t)\rangle\,,
\end{eqnarray}
where $\varepsilon_\alpha$ and $|\phi_\alpha(t)\rangle = |\phi_\alpha(\vec{r}_1,\vec{r}_2,t)\rangle$ are the eigenvalues and eigenstates of the Floquet Hamiltonian ${\cal H}_F = H - i\frac{\partial}{\partial t}$, 
referred to as \textit{quasienergies} and \textit{Floquet states}, respectively.

By expanding the Floquet states in a Fourier series,
\begin{equation}
|\phi_\alpha(t)\rangle = \sum_{k=-\infty}^\infty \ee^{-i k\omega t}\,|\phi_\alpha^k\rangle\,,
\end{equation}
the eigenvalue problem ${\cal H}_F|\phi_\alpha(t)\rangle = \varepsilon_\alpha|\phi_\alpha(t)\rangle$ reduces to a set of coupled equations for the Fourier components:
\begin{eqnarray}\nonumber
(H_0 +F_{\textrm{st}}(x_1+x_2)- k\omega)\,|\phi_\alpha^k\rangle \\ \nonumber
+ \frac{F}{2}(x_1+x_2)\left(|\phi_\alpha^{k+1}\rangle + |\phi_\alpha^{k-1}\rangle\right)\\ 
= \varepsilon_\alpha\,|\phi_\alpha^k\rangle\,.\label{ec:EVP-floquet}
\end{eqnarray}
Here, $k$ is the Floquet quantum number, which labels the number of photons exchanged with the field \cite{shirley1,delande8}.

We solve Equation (\ref{ec:EVP-floquet}) in the atomic basis
\begin{equation}\label{ec:AtomicBasis}
\{|\varphi_j^{L,k}\rangle\},\qquad |\varphi_j^{L,k}\rangle = |k\rangle \otimes |\varphi_j^L\rangle\,,
\end{equation}
where $|\varphi_j^L\rangle$ are the eigenstates of the unperturbed atom defined in (\ref{ec:TISE}), and $k$ is the Floquet index.
In this basis, the matrix form of (\ref{ec:EVP-floquet}) becomes
\begin{equation}\label{ec:matEGP1}
\left(\mathbf{h}_0 - k\omega\,\mathbbm{1} + \mathbf{F}_{\textrm{st}} + \mathbf{F}\right)\Phi_\alpha = \varepsilon_\alpha\,\Phi_\alpha\,,
\end{equation}
where $\Phi_\alpha$ denotes the vector representation of the Floquet state, and $\mathbf{h}_0$ is a diagonal matrix containing the field-free atomic eigenenergies.
The matrices $\mathbf{F}_{\textrm{st}}$ and $\mathbf{F}$ describe the dipole interaction terms associated with the static and periodic fields, respectively. 
Both are symmetric and exhibit block structure with couplings given by $\Delta k = \pm 1$ and $\Delta L = \pm 1$, reflecting the selection rules for dipole transitions.

The numerical implementation requires truncating the Floquet basis to a finite number of $k$-blocks and restricting the total angular momentum values:
\begin{equation}
k_{\textrm{min}} \leq k \leq k_{\textrm{max}}, \qquad L = 0, \dots, L_{\textrm{max}}\,.
\end{equation}

\subsection{Complex scaling method}
\label{subsec:complex-scaling}

Resonance states appearing in the spectrum of the unperturbed and driven helium atom are computed using the complex rotation (or complex scaling) method \cite{AC:CMP22-269, balslev1, S:AoM97-247, R:ARPC33-223, H:PRep99-1}.
This transformation is generated by the non-unitary operator
\begin{eqnarray}
R(\theta) = \exp\left(-\theta\,\frac{\vec{r}\cdot\vec{p} + \vec{p}\cdot\vec{r}}{2} \right)\,,
\end{eqnarray}
which acts on the position and momentum operators as $\vec{r} \to \vec{r}\,\ee^{\ii\theta}$ and $\vec{p} \to \vec{p}\,\ee^{-\ii\theta}$, with $\vec{r} = (\vec{r}_1, \vec{r}_2)$ and $\vec{p} = (\vec{p}_1, \vec{p}_2)$.

As a result, the complex-rotated Hamiltonian $H(\theta) = R(\theta) H R^{-1}(\theta)$ becomes non-Hermitian, and the wave functions associated with resonance states become square-integrable. 
The corresponding complex eigenvalues take the form $E - \ii\Gamma/2$, where $E$ is the resonance energy and $\Gamma$ is the decay rate, which is inversely related to the resonance lifetime.

For both field-free and driven helium, the complex-scaled Hamiltonian is diagonalized in a selected energy region using a Lanczos-based iterative method \cite{lanczos1,krug1}, 
and convergence is verified with respect to variations in the rotation angle $\theta$.

\subsection{Classical frozen planet configuration}
\label{subsec:frozen-planet-classical}

The classical FPC refers to a collinear and dynamically stable arrangement of the helium atom~\cite{Richter1992,Richter1990,schlagheck1}, in which the inner electron follows highly eccentric Keplerian orbits while the outer electron remains nearly ``frozen'' at an equilibrium distance from the nucleus.

\begin{figure}
\begin{center}
\includegraphics[width=0.26\textwidth]{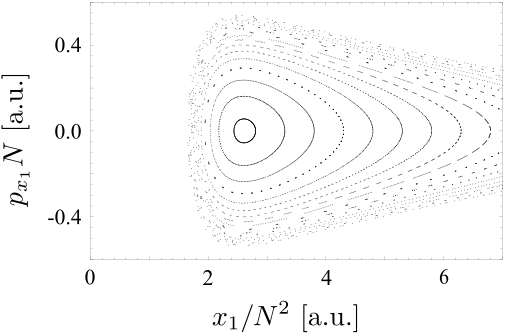}
\caption{Phase space of the outer electron in the collinear FPC, constructed from a Poincaré section taken when the inner electron reaches the nucleus.}
\label{fig:FPC}
\end{center}
\end{figure}

The phase space structure associated with the FPC is illustrated in Figure~\ref{fig:FPC}, through a Poincar\'e surface of section constructed by plotting the position $x_1$ and momentum $p_1$ of the 
outer electron each time the inner electron reaches the nucleus. This reveals a stable island of regular motion associated with the outer electron's slow oscillations around its equilibrium position.

Within the framework of adiabatic invariants \cite{lichtenberg83}, the fast Kepler motion of the inner electron generates an effective potential that confines the dynamics of the outer electron. 
From the shape of this potential, one can extract the intrinsic frequency $\omega_{\textrm{I}} = 0.3\,N^{-3}$ of small oscillations around the equilibrium position $x_{\textrm{min}} = 2.6\,N^2$, as well 
as the minimum static field strength required for ionization, $F_{\textrm{I}} = 0.03\,N^{-4}$, where $N$ is the action variable of the inner electron's motion \cite{ostrovsky1,schlagheck2}. 
In Section~\ref{subsec:fps-identification}, this quantity will be identified with the principal quantum number of the inner electron.

When an external monochromatic field is applied, the frozen planet dynamics evolves in a five-dimensional phase space, described by the coordinates and momenta of the two electrons, together with the phase $\omega t$ of the field.

For near-resonant driving, $\omega \approx \omega_{\textrm{I}}$, and moderate field amplitudes $F < F_{\textrm{I}}$, the hierarchy of time scales --- fast oscillations of the inner electron versus 
slow motion of the outer electron --- allows for a reduced two-dimensional representation of the dynamics using a two-step Poincar\'e section method \cite{schlagheck1,Schlagheck1999}.

The first section is constructed by recording the position and momentum of the outer electron $(x_1, p_1)$ each time the inner electron passes through the nucleus at $x_2 = 0$. 
These points are then interpolated to obtain a smooth trajectory that describes the outer electron’s slow motion. A second section is then taken at fixed phase of the driving 
field, $\omega t = \phi_0\ ({\textrm{mod}}\ 2\pi)$, to project the five-dimensional flow onto a two-dimensional surface.

\begin{figure}
\begin{center}
\includegraphics[width=0.48\textwidth]{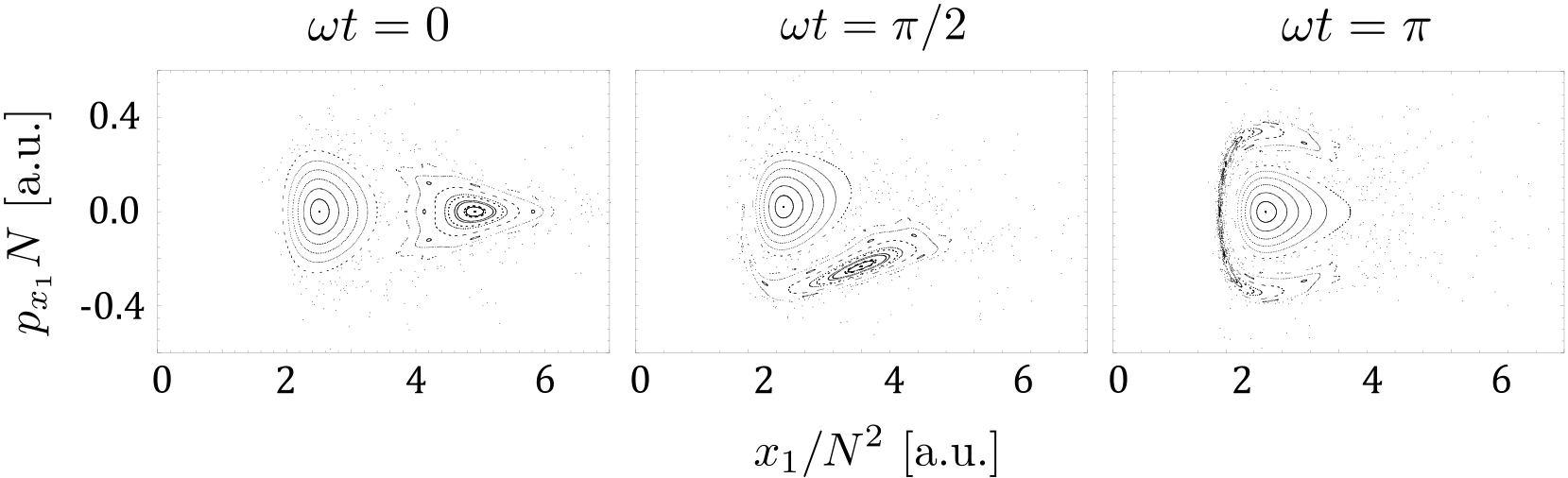}
\caption{Two-step Poincar\'e surface of section for the driven FPC, computed for field amplitude $F = 0.005N^{-4}\,\rm a.u.$ and frequency $\omega = 0.2N^{-3}\,\rm a.u.$, 
shown for different values of the field phase $\omega t$.}
\label{fig:DrivenFPC}
\end{center}
\end{figure}

Figure~\ref{fig:DrivenFPC} shows the resulting phase space for a driving field with frequency $\omega = 0.2N^{-3}\,\rm a.u.$ and amplitude $F = 0.005N^{-4}\,\rm a.u.$, 
for three representative values of the field phase $\omega t$. The dynamics is organized around two regular islands embedded in a chaotic sea: the intrinsic island, 
centered at $x_1 = 2.5N^2$, and the resonance island, centered at $x_1 = 4.9N^2$. The intrinsic island remains essentially unaffected by the driving, while the resonance island oscillates around it at the driving frequency.

Small deviations from the collinear configuration—such as transverse displacements of the outer electron—can destabilize the correlated frozen planet motion \cite{Schlagheck1999}. 
In the presence of a periodic driving field, these perturbations are amplified by the torque exerted on the electronic distribution, resulting in a progressive breakdown of the correlated dynamics. 
After a few field cycles, the configuration may rotate chaotically, leading to strong electron-electron interactions and eventual ionization.

The application of a weak static electric field along the polarization axis can suppress this destabilization mechanism by limiting transverse excursions and restoring collinear motion~\cite{ostrovsky1}.
The static component introduces a restoring force that limits large-angle deviations, thereby confining the dynamics within regular regions of the full phase space. 
This classical stabilization mechanism provides the theoretical motivation for the combined use of periodic and static fields discussed in the following sections.

\section{Results}
\label{sec:results}

\subsection{Identification of frozen planet states}
\label{subsec:fps-identification}

Frozen planet states represent a class of highly correlated doubly excited states in helium that emerge as quantum manifestations of the classical FPC \cite{schlagheck2,madronero:epl05,madronero08:pra,richter6,eichmann1,heber1,camus1,percival1}. These states are characterized by small decay rates compared to other eigenstates, and are organized in series converging to single ionization thresholds $I_N$ (starting from $N=3$), where $N$ corresponds to the principal quantum number of the inner electron. 
We denote these states by ${\cal F}^N_{n_{\textrm{F}}}$, where $n_{\textrm{F}}$ labels their position in the corresponding series.

Tables~\ref{tab:FPS_1S} and~\ref{tab:FPS_1P} report the real and imaginary parts of the complex energy, as well as the expectation value of $\langle\cos\theta_{12}\rangle$, for the lowest four frozen planet states in the singlet sector, with total angular momenta $L=0,1$ and $N = 3,\dots,7$.
Since FPS are localized along a collinear configuration, the value of $\langle\cos\theta_{12}\rangle$ is expected to approach unity. 
This localization becomes more pronounced with increasing $N$, consistent with the semiclassical nature of these highly excited correlated states.

A complete picture of FPS localization is provided by analyzing their probability densities both in configuration space and in phase space. Due to the rotational invariance of the system, the conditional probability density 
at a given $\theta_{12}$ depends only on the relative position of the electrons. Thus, for $\theta_{12}=0$, it is possible to compute the density by aligning both electrons along the $x$-axis, setting $r_1 = x_1$, $r_2 = x_2$, with $x_1, x_2 > 0$ and $y_1 = y_2 = 0$.

\begin{figure}
\begin{center}
\includegraphics[width=0.32\textwidth]{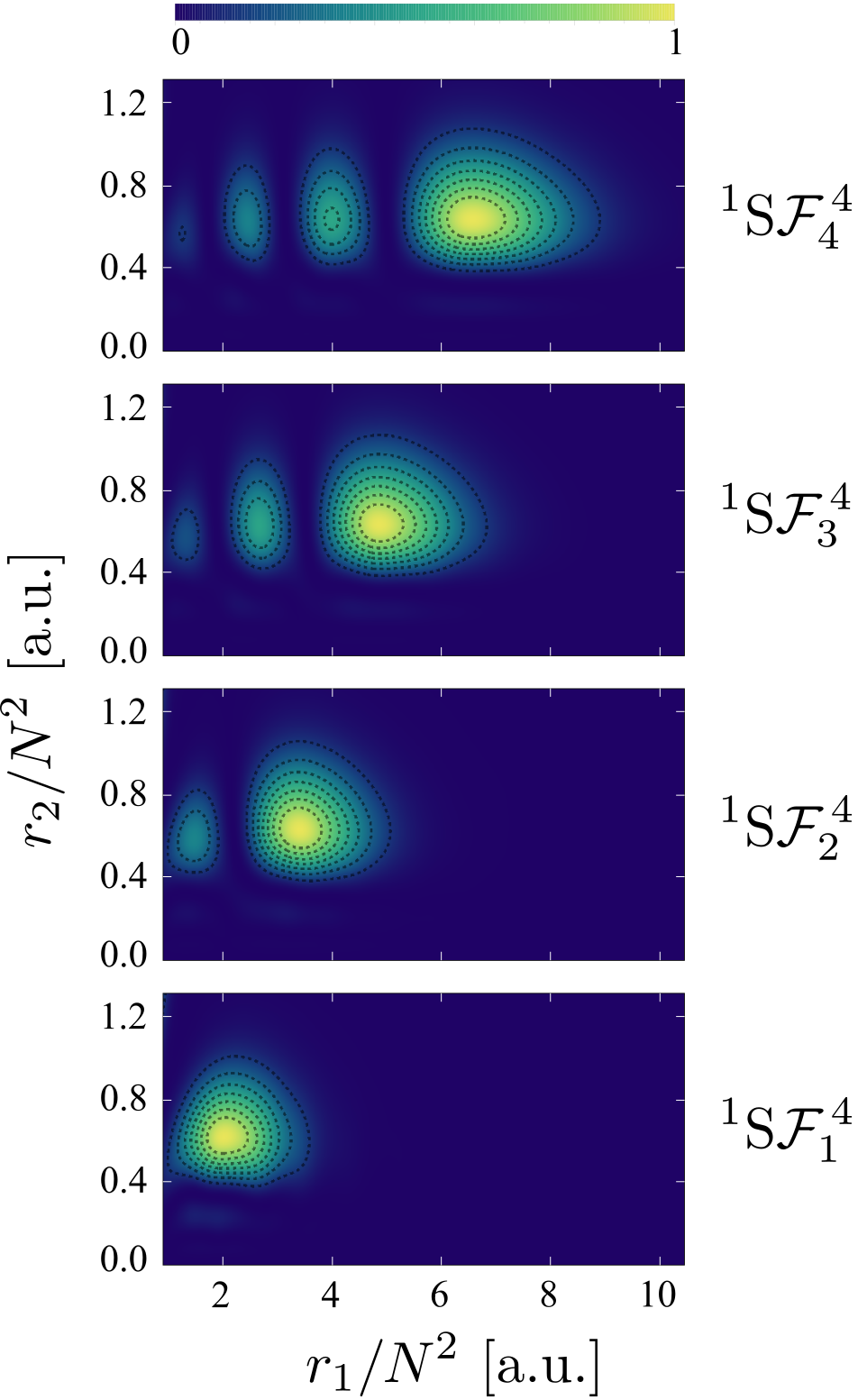}
\caption{(Color online) Conditional probability density ($\theta_{12}=0$) of the lowest $\isotope[1]{S}$ frozen planet states in the $N=4$ series. 
For the ground state, the maximum of the probability density is localized near the equilibrium position of the outer electron. 
For excited FPS, the distribution shifts outward along the $r_1$ axis while remaining centered in $r_2$, consistent with the structure of the classical configuration.}
\label{fig:ConditionalFPS}
\end{center}
\end{figure}

In Figure~\ref{fig:ConditionalFPS}, we display the conditional density of the lowest $\isotope[1]{S}$ FPS below the $N=4$ ionization threshold. 
The inner electron is localized near the nucleus, while the outer electron exhibits a broader distribution. 
For the ground FPS, the density maximum lies near the classical equilibrium position $x_{\textrm{min}}$ of the outer electron in the FPC. 
For excited FPS, the density along the inner electron axis remains centered near the nucleus, whereas the maximum along the outer electron axis increases with $n_{\textrm{F}}$, reflecting higher classical amplitudes.

\begin{table*}
\begin{center}
{\footnotesize
\setlength{\arraycolsep}{7pt}
\renewcommand{\arraystretch}{1.2}
$
\begin{array}{cccccccccccc}
\hline\hline
\rule[-1.8ex]{0pt}{4.8ex}
& & \multicolumn{3}{c}{\textrm{This work}} & & \multicolumn{3}{c}{\textrm{Richter}~\textit{et al.}~\text{\cite{richter6}}} & & \multicolumn{2}{c}{\textrm{Burgers}~\textit{et al.}~\text{\cite{burgers1995}}} \\
\cline{3-5} \cline{7-9} \cline{11-12}
\rule[-1.8ex]{0pt}{4.8ex}
N & n_{\textrm{F}} & -E\,[\textrm{a.u.}] & \Gamma/2\,[\textrm{a.u.}] & \langle\cos\theta_{12}\rangle &
& -E\,[\textrm{a.u.}] & \Gamma/2\,[\textrm{a.u.}] & \langle\cos\theta_{12}\rangle &
& -E\,[\textrm{a.u.}] & \Gamma/2\,[\textrm{a.u.}] \\ 
\hline
\rule[-1ex]{0pt}{3.5ex} 
3 & 1 & 0.257\,370\,27 & 0.000\,007\,12 & 0.352 & & & & & & 0.257\,371\,610 & 0.000\,010\,564 \\
  & 2 & 0.244\,323\,59 & 0.000\,020\,27 & 0.448 & & & & & & 0.244\,324\,739 & 0.000\,021\,400 \\
  & 3 & 0.237\,310\,39 & 0.000\,016\,42 & 0.469 & & & & & & 0.237\,311\,202 & 0.000\,017\,021 \\
  & 4 & 0.233\,173\,07 & 0.000\,011\,94 & 0.484 & & & & & & 0.233\,173\,689 & 0.000\,012\,347 \\[1.2ex]

4 & 1 & 0.141\,067\,41 & 0.000\,010\,79 & 0.534 & & 0.141\,064\,16 & 0.000\,011\,74 & 0.534 & & 0.141\,064\,156 & 0.000\,011\,739 \\
  & 2 & 0.137\,091\,40 & 0.000\,001\,77 & 0.552 & & 0.137\,088\,22 & 0.000\,002\,48 & 0.552 & & 0.137\,088\,229 & 0.000\,002\,490 \\
  & 3 & 0.134\,230\,69 & 0.000\,002\,29 & 0.573 & & & & & & 0.134\,228\,598 & 0.000\,002\,711 \\
  & 4 & 0.132\,214\,01 & 0.000\,003\,04 & 0.584 & & & & & & 0.132\,212\,660 & 0.000\,003\,293 \\[1.2ex]

5 & 1 & 0.089\,571\,26 & 0.000\,002\,07 & 0.712 & & 0.089\,570\,80 & 0.000\,002\,02 & 0.712 & & & \\
  & 2 & 0.087\,561\,17 & 0.000\,006\,12 & 0.540 & & 0.087\,559\,62 & 0.000\,006\,60 & 0.521 & & & \\
  & 3 & 0.086\,099\,22 & 0.000\,000\,60 & 0.608 & & 0.086\,097\,67 & 0.000\,000\,79 & 0.609& & & \\
  & 4 & 0.085\,007\,53 & 0.000\,000\,53 & 0.607 & &  &  & & & & \\[1.2ex]

6 & 1 & 0.062\,053\,80 & 0.000\,000\,54 & 0.747 & & 0.062\,053\,56 & 0.000\,000\,56 & 0.747 & & & \\
  & 2 & 0.060\,840\,80 & 0.000\,000\,92 & 0.740 & & 0.060\,840\,21 & 0.000\,000\,98 & 0.737 & & & \\
  & 3 & 0.059\,938\,86 & 0.000\,001\,26 & 0.749 & & 0.059\,938\,22 & 0.000\,001\,45 & 0.743 & & & \\
  & 4 & 0.059\,247\,33 & 0.000\,000\,92 & 0.783 & &  &  & & & & \\[1.2ex]

7 & 1 & 0.045\,539\,29 & 0.000\,000\,07 & 0.780 & & 0.045\,538\,67 & 0.000\,000\,20 & 0.776& & & \\
  & 2 & 0.044\,762\,80 & 0.000\,001\,44 & 0.782 & & 0.044\,758\,56 & 0.000\,000\,36 & 0.763& & & \\
  & 3 & 0.044\,172\,97 & 0.000\,005\,89 & 0.656 & & 0.044\,161\,29 & 0.000\,006\,50 & 0.747 & & & \\
  & 4 & 0.043\,697\,51 & 0.000\,003\,39 & 0.701 & &  &  & & & & \\
\hline\hline
\end{array}
$
}
\end{center}
\caption{Energies, decay rates, and expectation values of $\langle\cos\theta_{12}\rangle$ for the four lowest $\isotope[1]{S}$ frozen planet states with $N = 3,\dots,7$. 
Our results (left columns) are compared with data from Refs.~\cite{richter6} and \cite{burgers1995}. 
The expectation values of $\langle\cos\theta_{12}\rangle$ are only available in Ref.~\cite{richter6}.}
\label{tab:FPS_1S}
\end{table*}

\begin{table*}
\begin{center}
{\footnotesize
\setlength{\arraycolsep}{7pt}
\renewcommand{\arraystretch}{1.2}
$
\begin{array}{cccccccc}
\hline\hline
\rule[-1.8ex]{0pt}{4.8ex}
& & \multicolumn{3}{c}{\textrm{This work}} & & \multicolumn{2}{c}{\textrm{Rost}~\textit{et al.}~\text{\cite{rost1997}}}  \\
\cline{3-5} \cline{7-8}
\rule[-1.8ex]{0pt}{4.8ex}
N & n_{\textrm{F}} & -E\,[\textrm{a.u.}] & \Gamma/2\,[\textrm{a.u.}] & \langle\cos\theta_{12}\rangle &
& -E\,[\textrm{a.u.}] & \Gamma/2\,[\textrm{a.u.}] \\ 
\hline
\rule[-1ex]{0pt}{3.5ex}
3 & 1 & 0.245\,517\,49 & 0.000\,000\,04 & 0.496 & & 0.245\,517\,652 & 0.000\,000\,068 \\
  & 2 & 0.238\,061\,59 & 0.000\,000\,08 & 0.515 & & 0.238\,061\,744& 0.000\,000\,106 \\
  & 3 & 0.233\,663\,16 & 0.000\,000\,08 & 0.520 & & 0.233\,663\,277& 0.000\,000\,102 \\
  & 4 & 0.230\,864\,50 & 0.000\,000\,07 & 0.520 & & 0.230\,864\,587& 0.000\,000\,086 \\[1.2ex]

4 & 1 & 0.138\,613\,88 & 0.000\,000\,40 &	0.600 & & 0.138\,613\,869& 0.000\,000\,412 \\
  & 2 & 0.135\,102\,26 & 0.000\,000\,14 &	0.617 & & 0.135\,102\,260& 0.000\,000\,154 \\
  & 3 & 0.132\,774\,21 & 0.000\,000\,04 &	0.625 & & 0.132\,774\,213& 0.000\,000\,059 \\
  & 4 & 0.131\,159\,88 & 0.000\,000\,02 &	0.628 & & 0.131\,159\,886& 0.000\,000\,032 \\[1.2ex]

5 & 1 & 0.088\,845\,26 & 0.000\,002\,91 &	0.672 & & 0.088\,845\,250& 0.000\,002\,921 \\
  & 2 & 0.086\,939\,07 & 0.000\,001\,70 &	0.694 & & 0.086\,939\,055& 0.000\,001\,712 \\
  & 3 & 0.085\,578\,31 & 0.000\,001\,31 &	0.699 & & 0.085\,578\,293& 0.000\,001\,323 \\
  & 4 & 0.084\,576\,54 & 0.000\,000\,46 &	0.696 & & 0.084\,576\,520& 0.000\,000\,470 \\[1.2ex]

6 & 1 & 0.061\,734\,15 & 0.000\,007\,67 &	0.735 & & 0.061\,734\,162& 0.000\,007\,671 \\
  & 2 & 0.060\,593\,31 & 0.000\,006\,37 &	0.757 & & 0.060\,593\,300& 0.000\,006\,363 \\
  & 3 & 0.059\,735\,22 & 0.000\,004\,05 &	0.756 & & 0.059\,735\,205 & 0.000\,004\,050 \\
  & 4 & 0.059\,075\,38 & 0.000\,003\,08 &	0.759 & & 0.059\,075\,368& 0.000\,003\,092 \\[1.2ex]

7 & 1 & 0.045\,366\,34 & 0.000\,002\,76 &	0.728 & & 0.045\,366\,336& 0.000\,002\,763 \\
  & 2 & 0.044\,626\,42 & 0.000\,004\,31 &	0.721 & & 0.044\,626\,424& 0.000\,004\,312 \\
  & 3 & 0.044\,056\,63 & 0.000\,003\,57 &	0.759 & & 0.044\,056\,650& 0.000\,003\,576 \\
  & 4 & 0.043\,601\,08 & 0.000\,002\,57 &	0.770 & & 0.043\,601\,101& 0.000\,002\,570 \\
\hline\hline
\end{array}
$
}
\end{center}
\caption{Energies, decay rates, and expectation values of $\langle\cos\theta_{12}\rangle$ for the four lowest $\isotope[1]{P}$ frozen planet states with $N = 3,\dots,7$. 
Our results (left columns) are compared with energy and decay rate data from Ref.~\cite{rost1997}. 
The expectation values of $\langle\cos\theta_{12}\rangle$ are not reported in that reference.}
\label{tab:FPS_1P}
\end{table*}

\begin{figure}
\begin{center}
\includegraphics[width=0.3\textwidth]{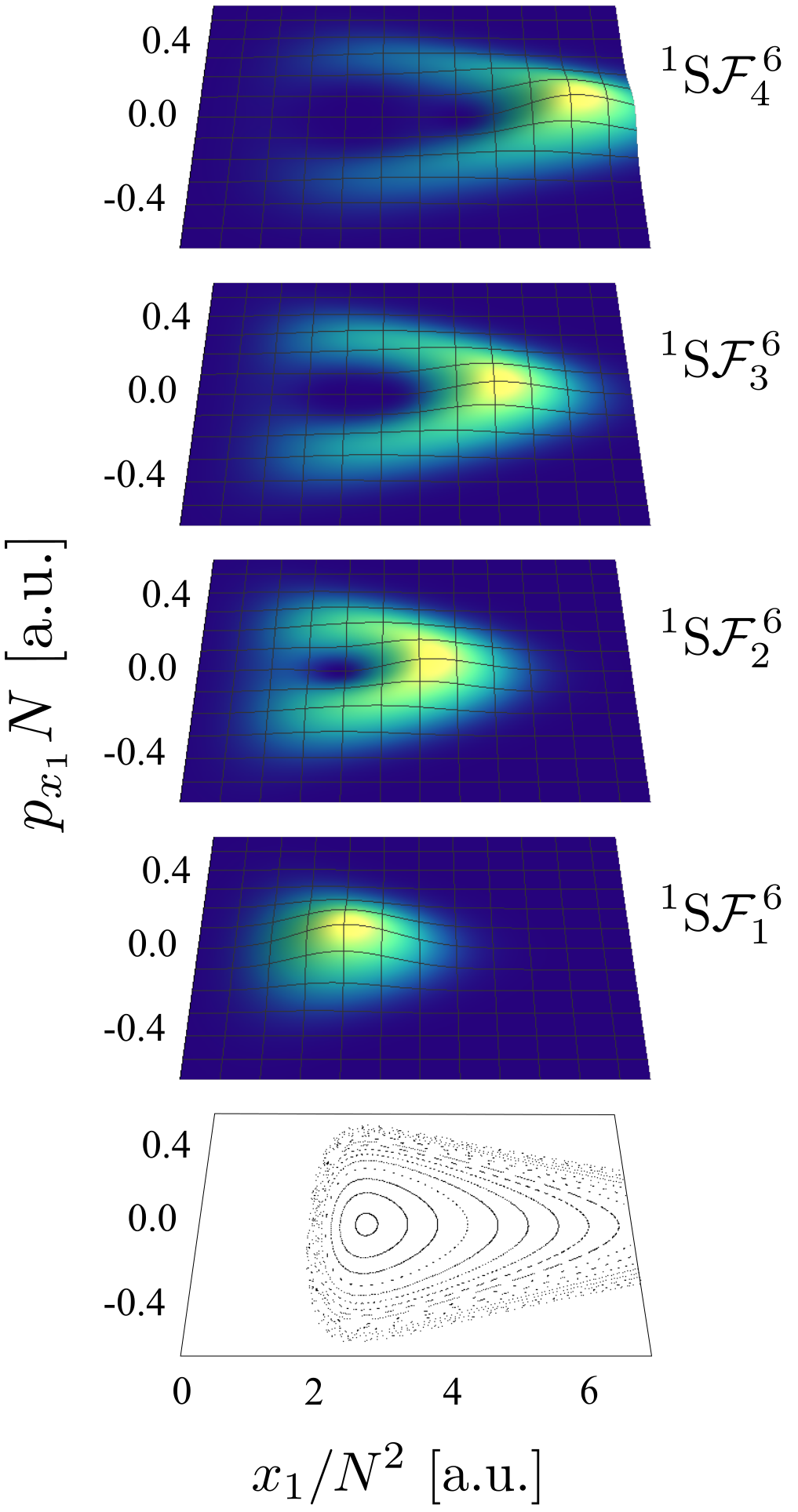}
\caption{(Color online) Phase space Husimi distributions of the $\isotope[1]{S}$ frozen planet states in the $N=6$ series. 
The ground state $\isotope[1]{S}{\cal F}^{\,6}_1$ is localized near the equilibrium position of the outer electron. 
Excited FPS exhibit localization around higher-energy periodic orbits of the classical configuration.}
\label{fig:HusimiFPS}
\end{center}
\end{figure}

Figure~\ref{fig:HusimiFPS} shows the Husimi distributions of the first four $\isotope[1]{S}$ frozen planet states in the $N=6$ series. 
These distributions are projected onto the phase space of the outer electron and compared with the classical Poincaré section of the FPC (bottom panel). 
The ground state ${\cal F}^{\,6}_1$ exhibits a clear localization near the equilibrium position of the outer electron in the classical phase space, indicating that the wave function 
is concentrated around the stable fixed point of the FPC. As the excitation number $n_{\textrm{F}}$ increases, the FPS become localized along classical periodic orbits of higher energy within the same phase space structure. 
This correspondence confirms that FPS form quantized families of two-electron states whose structure and dynamics are governed by the underlying classical configuration.

\subsection{Frozen planet states under periodic driving}
\label{subsec:fps-driven}

The direct numerical treatment of the full three-dimensional helium atom subjected to a periodic external field represents a significant computational challenge. 
This has historically limited the exploration of nondispersive wave packets beyond reduced-dimensionality models \cite{schlagheck2,madronero08:pra}. 
However, it has been shown that such wave packets can be efficiently characterized by solving the TDSE in the atomic basis, i.e., 
the basis in which the unperturbed Hamiltonian $H_0$ is diagonal \cite{GonzalezMelan2020}. 
Of particular relevance is the observation that NDWP in helium emerge from a near-resonant coupling between two frozen planet states. 

Motivated by this mechanism, we analyze the Floquet spectrum of the periodically driven helium atom in the energy region below the $N=6$ ionization threshold, 
focusing on the coupling between the singlet states $\isotope[1]{S}{\cal F}^{\,6}_2$ and $\isotope[1]{P}{\cal F}^{\,6}_1$.

The field parameters are selected such that the driving frequency is approximately resonant with the energy separation between these two 
FPS, i.e., $\omega \approx |E(\isotope[1]{S}{\cal F}^{\,6}_2) - E(\isotope[1]{P}{\cal F}^{\,6}_1)|$, and the field amplitude satisfies $F < F_{\textrm{I}}$. 
The Floquet eigenvalue problem (\ref{ec:matEGP1}) is solved in an atomic basis including 13 Floquet blocks ($k_{\textrm{min}} = -6$ to $k_{\textrm{max}} = 6$) and total angular momenta $L = 0, \dots, 5$. 
This choice of basis size is guided by convergence tests, which show that increasing the number of Floquet blocks or angular momentum channels does not significantly alter the results presented here.
Similar behavior was observed in reduced-dimensional helium~\cite{GonzalezMelan2020}, where nondispersive wave packets emerge from the interaction of a small number of near-resonant states.

Among the resulting Floquet eigenstates, we identify those dominated by a single frozen planet state ${\cal F}^N_j$ through the maximal overlap $|\langle{\cal F}^N_j|\psi\rangle|^2$, 
and denote the corresponding dressed states as ${\cal W}^N_j$.

\begin{figure}
\begin{center}
\includegraphics[width=0.48\textwidth]{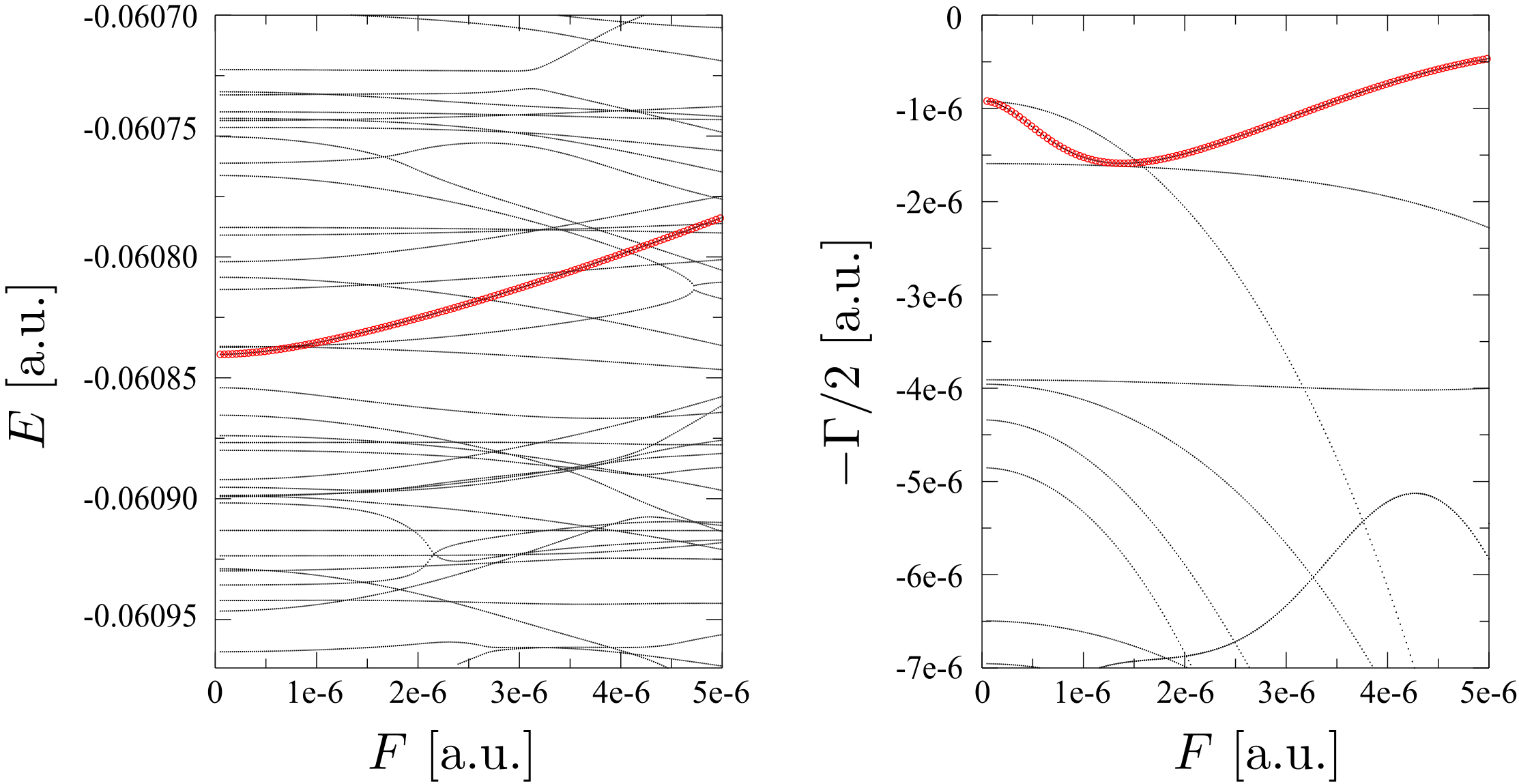}
\end{center}
\caption{Floquet quasienergies $E$ (left) and decay rates $\Gamma$ (right) as functions of the driving amplitude $F$, for fixed frequency $\omega = 0.0008\,\rm a.u.$ The Floquet state ${\cal W}^6_2$ is highlighted in red.}
\label{fig:Energias}
\end{figure}

Figure~\ref{fig:Energias} shows the real and imaginary parts of the Floquet quasienergies as functions of the field amplitude for a fixed driving frequency $\omega = 0.00088\,\rm a.u.$ One of the eigenstates, 
labeled ${\cal W}^6_2$ (highlighted in red), maintains the largest overlap with the unperturbed FPS $\isotope[1]{S}{\cal F}^{\,6}_2$ across a finite interval of field strengths. 
The smooth variation of its quasienergy, together with a relatively small decay rate over that interval, points to the formation of a long-lived state.

\begin{figure}
\begin{center}
\includegraphics[width=0.48\textwidth]{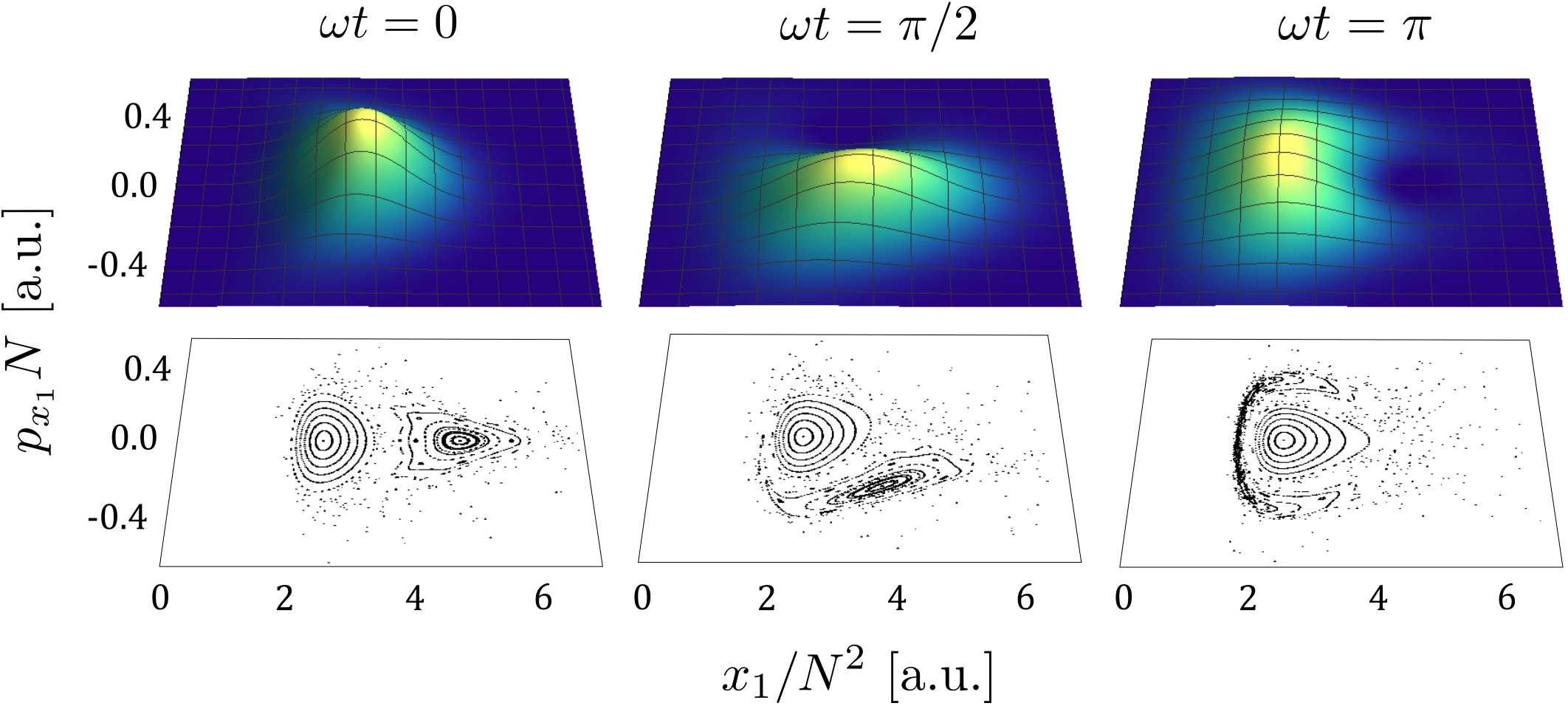}
\end{center}
\caption{Husimi distributions of the Floquet state ${\cal W}^6_2$ at different field phases (top), computed for $F=3.9\times10^{-6}\,\rm a.u.$ and $\omega=0.00088\,\rm a.u.$ 
The bottom panel shows the corresponding classical resonance island in the driven FPC.}
\label{fig:HusimiNDWP}
\end{figure}

Figure~\ref{fig:HusimiNDWP} displays the Husimi distributions of this Floquet state. These distributions remain well localized around the resonance island for different phases of the driving 
field, following the classical 1:1 resonance trajectory. This localization is a direct manifestation of a nondispersive wave packet: a quantum state that follows a classical periodic orbit without significant spreading.

\begin{figure}
\begin{center}
\includegraphics[width=0.48\textwidth]{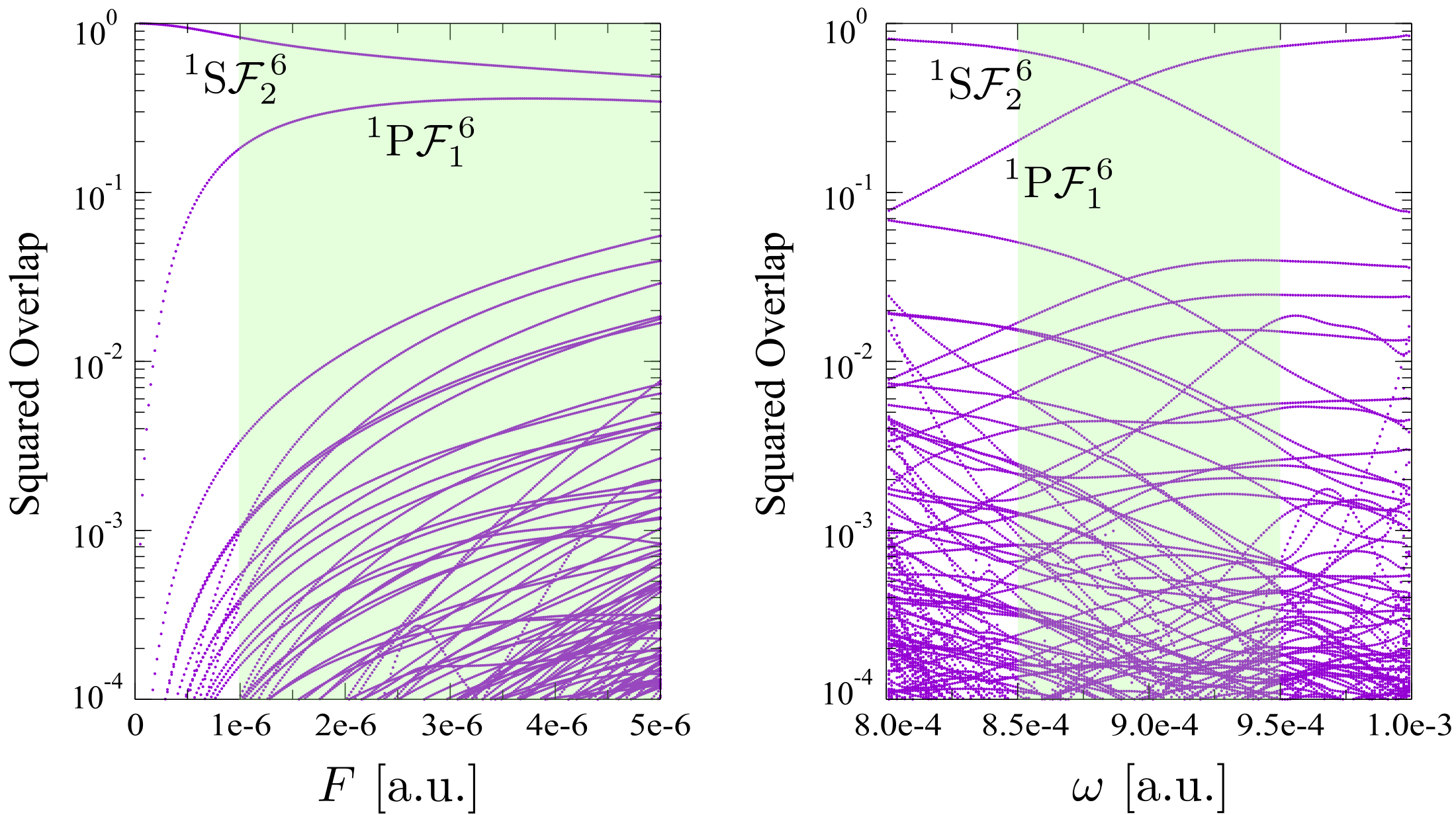}
\end{center}
\caption{Overlaps between the Floquet state ${\cal W}^6_2$ and the elements in the basis. 
Left: overlap vs. field amplitude $F$ at fixed frequency $\omega = 0.00088\,\rm a.u.$ 
Right: overlap vs. frequency $\omega$ at fixed amplitude $F = 3.9\times10^{-6}\,\rm a.u.$ 
Green shaded regions indicate the stability window where ${\cal W}^6_2$ exhibits nondispersive behavior.}
\label{fig:Overlaps}
\end{figure}

Further insight into the structure and stability of the nondispersive wave packet ${\cal W}^6_2$ is provided by the analysis of overlaps shown in Figure~\ref{fig:Overlaps}. 
The left panel shows the overlaps as a function of the field amplitude $F$ at fixed frequency $\omega = 0.00088\,\rm a.u.$, while the right panel displays the dependence 
on $\omega$ at fixed $F = 3.9 \times 10^{-6}\,\rm a.u.$ In both cases, ${\cal W}^6_2$ arises as a coherent superposition of the FPS states $\isotope[1]{S}{\cal F}^{\,6}_2$ and $\isotope[1]{P}{\cal F}^{\,6}_1$, 
with the dominant character switching across the resonance. All other contributions are negligible on a logarithmic scale.

A systematic sweep of the parameter space confirms that ${\cal W}^6_2$ survives as a robust nondispersive state for driving frequencies in the range
$0.00085\ \text{a.u.}\lesssim\omega\lesssim0.00095\ \text{a.u.}$ at fixed $F = 3.9\times10^{-6}\,\rm a.u.$, and for amplitudes
$1.0\times10^{-6}\ \text{a.u.}\lesssim F\lesssim5.0\times10^{-6}\,\rm a.u.$\ at fixed $\omega = 0.00088\,\rm a.u.$
Throughout this two-parameter window, the Floquet state ${\cal W}^6_2$ remains well localized on the $1{:}1$ resonance island, as evidenced by its Husimi distribution (Figure~\ref{fig:HusimiNDWP}) and 
the high overlap with only two dominant FPS basis states. 
This confirms the robustness of ${\cal W}^6_2$ as a nondispersive wave packet under purely periodic driving.

\begin{figure}
\begin{center}
\includegraphics[width=0.48\textwidth]{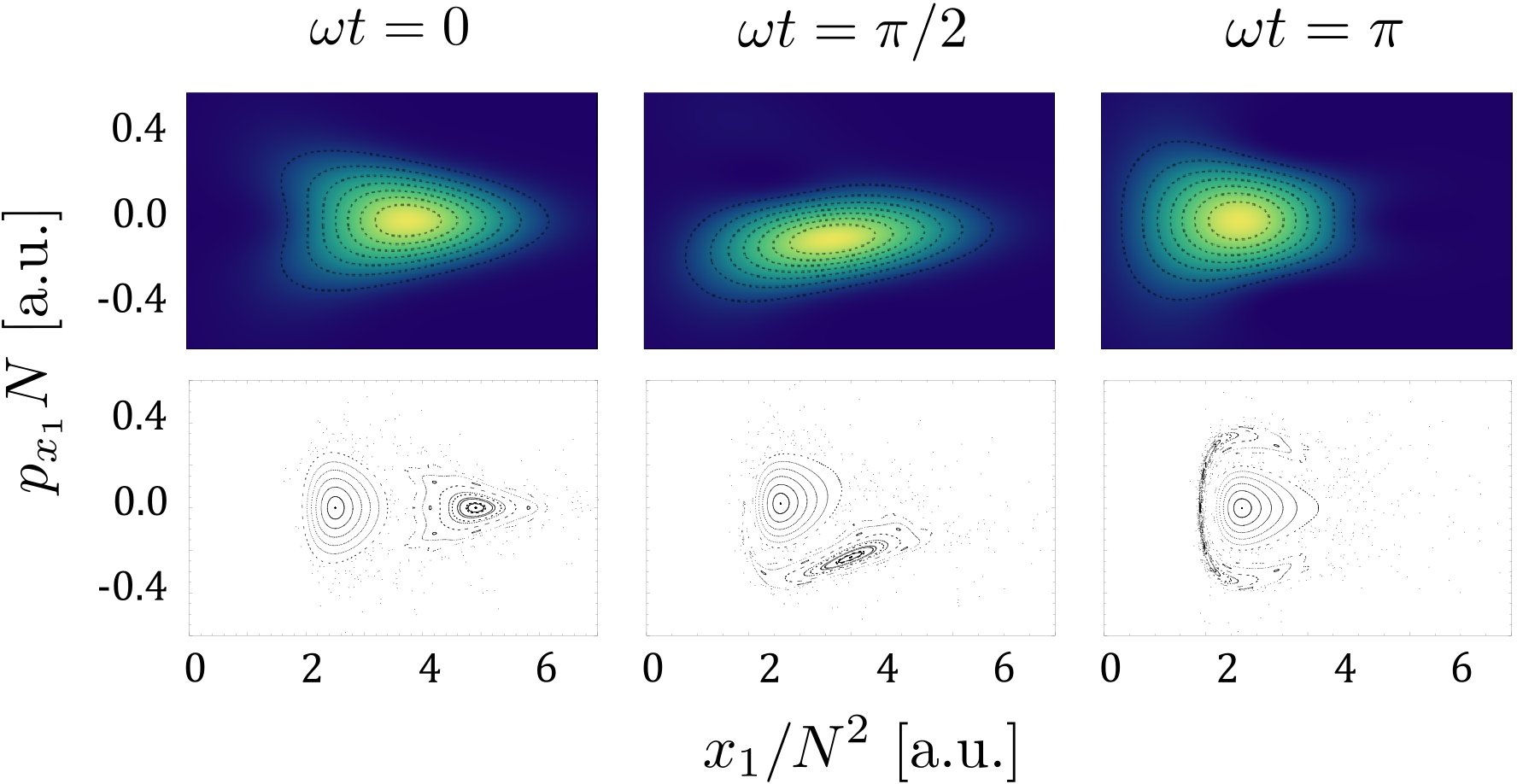}
\end{center}
\caption{Husimi distributions of a Floquet state below the $N=5$ ionization threshold, computed at $F = 8.0\times10^{-6}\,\rm a.u.$ and $\omega = 0.0013\,\rm a.u.$ 
Top panels show the distribution at different phases of the driving field, revealing localization on the classical $1{:}1$ resonance island. 
The bottom panel depicts the corresponding classical phase space structure for the driven FPC.}
\label{fig:HusimiNDWP_N5}
\end{figure}

Although our numerical exploration has focused on a Floquet state in the $N=6$ series, similar behavior is also observed in other thresholds.
Figure~\ref{fig:HusimiNDWP_N5}, for instance, shows the Husimi distribution of a Floquet state associated with the $N=5$ series, which exhibits localization on the classical 
resonance island analogous to that of ${\cal W}^6_2$.

In what follows, we explore the effect of a static field component on the stability of these driven wave packets.

\subsection{Decay behavior of frozen planet states in a static field}
\label{subsec:fps-static-field}

We now analyze the effect of a weak static electric field on the stability of selected frozen planet states in the absence of periodic driving.
This configuration allows us to isolate the role of the static component and to examine its influence on the decay properties of the FPS involved in the formation of the nondispersive wave packet.

\begin{figure}
\begin{center}
\includegraphics[width=0.48\textwidth]{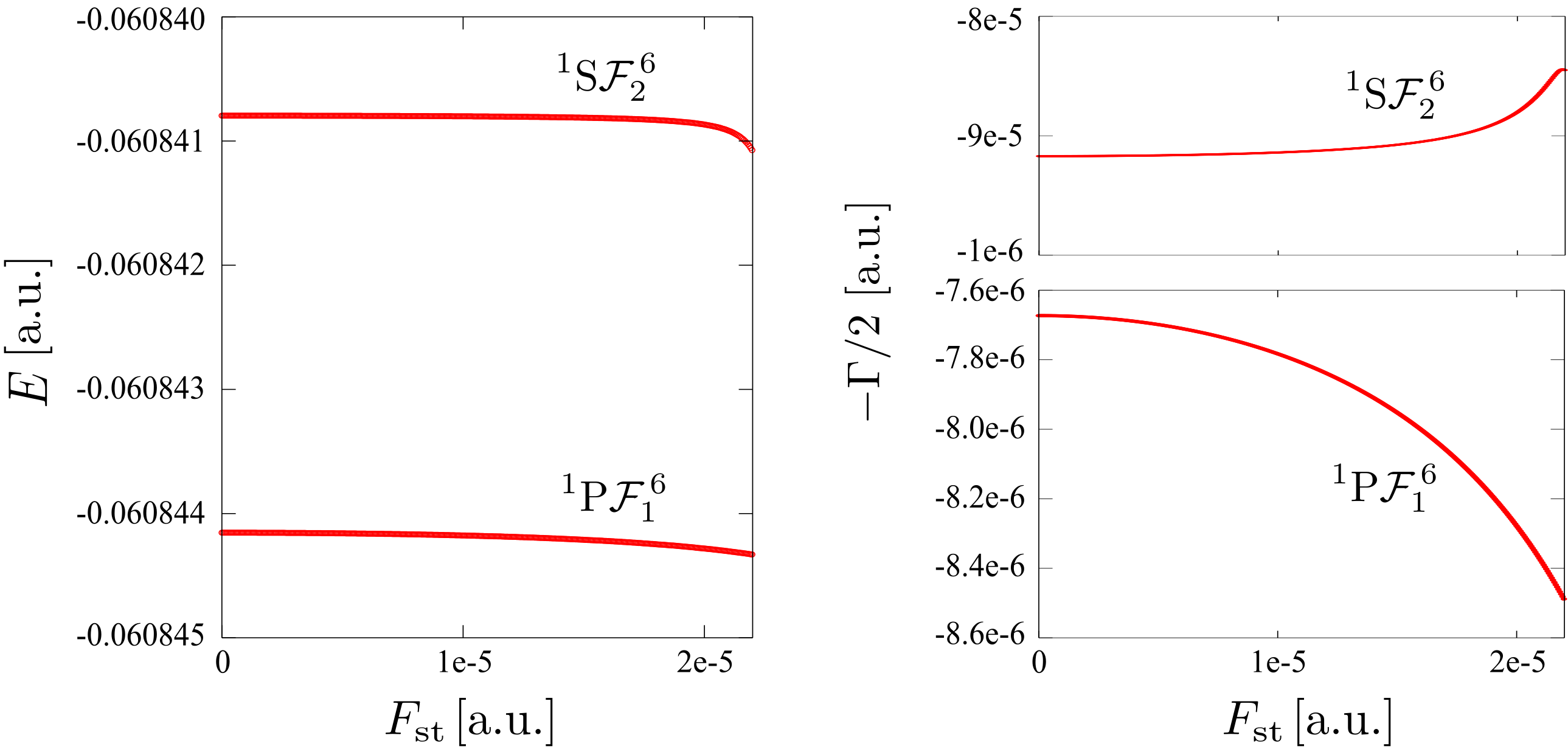}
\end{center}
\caption{Energies (left) and decay rates (right) of the frozen planet states  $\isotope[1]{S}{\cal F}^{\,6}_2$ and $\isotope[1]{P}{\cal F}^{\,6}_1$ as a function of the static electric field.
This range extends up to the value for the minimum static field necessary to ionize the configuration, given by $F_{\textrm{I}} = 0.03 N^{-4} \approx 2.3\times10^{-5}\,\rm a.u.$
}
\label{fig:fps_staticfield}
\end{figure}

Specifically, we focus on the states $\isotope[1]{S}{\cal F}^{\,6}_2$ and $\isotope[1]{P}{\cal F}^{\,6}_1$, which—according to the analysis presented in Section~\ref{subsec:fps-driven}—constitute 
the dominant components of the Floquet state ${\cal W}^6_2$.
Figure~\ref{fig:fps_staticfield} shows the dependence of their complex energies on the static field strength $F_{\textrm{st}}$.

For the state $\isotope[1]{S}{\cal F}^{\,6}_2$, the decay rate $\Gamma$ increases gradually with $F_{\textrm{st}}$ up to approximately $1.5 \times 10^{-5}\,\rm a.u.$, beyond which it grows more rapidly and 
reaches a maximum near $2.23 \times 10^{-5}\,\rm a.u.$—the upper limit of the explored range.
In contrast, the decay rate of $\isotope[1]{P}{\cal F}^{\,6}_1$ decreases monotonically across the entire interval.

Importantly, the real parts of the energies remain nearly constant throughout the interval, ensuring that the resonance condition $\omega \approx |E(\isotope[1]{S}{\cal F}^{\,6}_2) - E(\isotope[1]{P}{\cal F}^{\,6}_1)|$ remains valid.
This condition will be crucial in the following section, where the effects of combined periodic and static fields are analyzed.

\subsection{Effect of a static field on periodically driven nondispersive wave packets}
\label{subsec:ndwp-static-field}

We now consider the combined action of a weak periodic drive and a collinear static electric field.
Our analysis concentrates on the Floquet state ${\cal W}^6_2$, the nondispersive wave packet identified in Section~\ref{subsec:fps-driven}.
Given that ${\cal W}^6_2$ is primarily composed of the FPS $\isotope[1]{S}{\cal F}^{\,6}_2$, its decay rate is expected to mirror the behavior observed in Section~\ref{subsec:fps-static-field}, suggesting the possibility for lifetime enhancement.

\begin{figure}
\begin{center}
\includegraphics[width=0.36\textwidth]{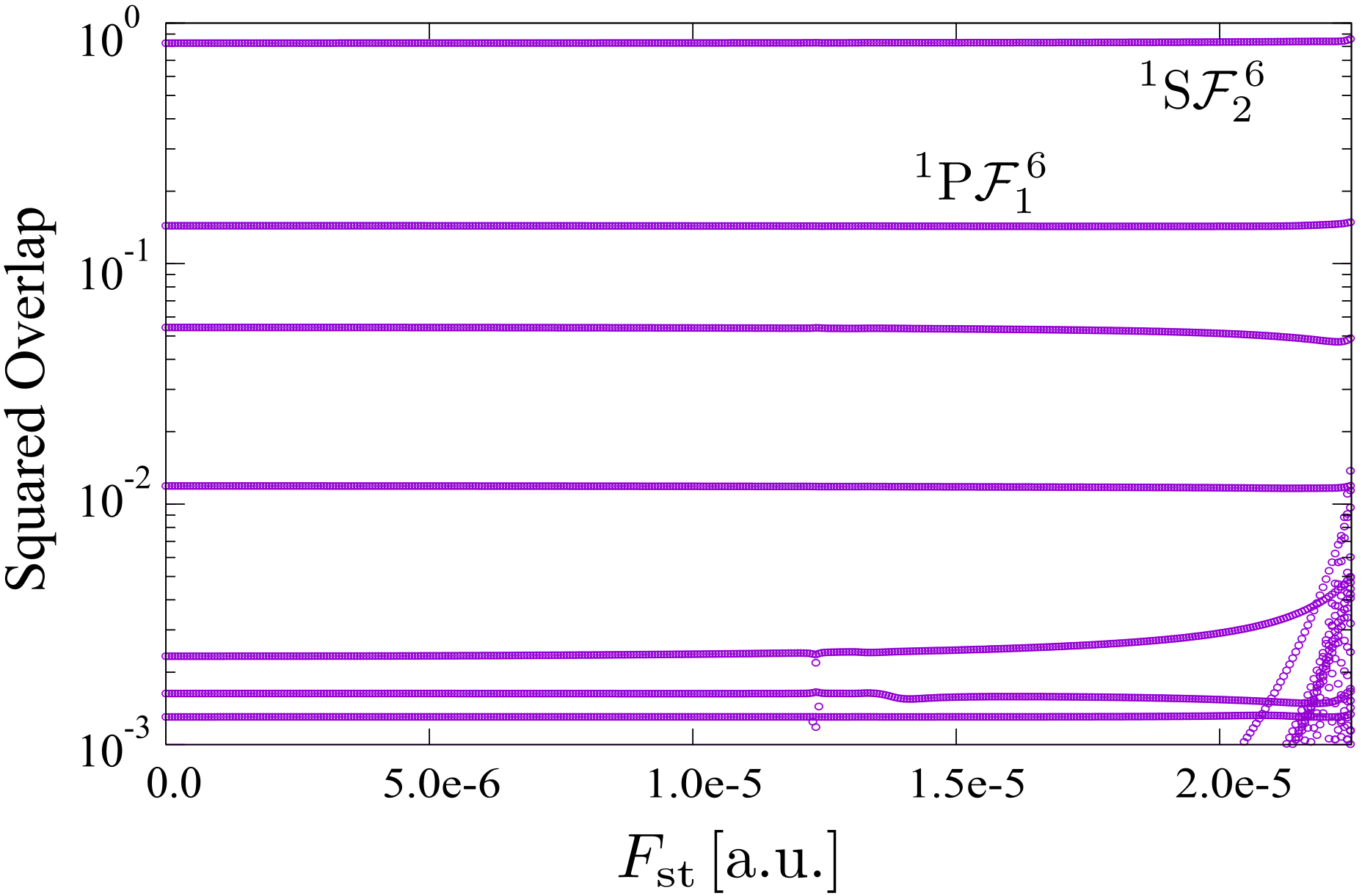}
\end{center}
\caption{Overlaps of the wave packet ${\cal W}^6_2$ with the basis states, as a function of the static electric field at fixed periodic driving with 
amplitude $ F= 1.0 \times 10^{-6} \,\rm a.u.$  and frequency $ \omega= 0.00089 \,\rm a.u.$
}
\label{fig:overlapsStaticField}
\end{figure}

Figure~\ref{fig:overlapsStaticField} shows the overlaps $|\langle{\cal F}^{\,6}_j|\psi\rangle|^2$ between ${\cal W}^6_2$ and the FPS basis states as $F_{\textrm{st}}$ is swept from $0$ to $2.3 \times 10^{-5}\,\rm a.u.$, 
with $\omega = 0.00089\,\rm a.u.$\ and $F = 1.0 \times 10^{-6}\,\rm a.u.$\ fixed.
Throughout this interval, the wave packet composition remains essentially constant, with $\sim 84\%$ $\isotope[1]{S}{\cal F}^{\,6}_2$ and $\sim 15\%$ $\isotope[1]{P}{\cal F}^{\,6}_1$, while the Husimi 
distributions remain sharply localized on the classical $1{:}1$ resonance island as in Figure~\ref{fig:HusimiNDWP}, confirming the nondispersive character of the state.

\begin{figure}
\begin{center}
	\includegraphics[width=0.48\textwidth]{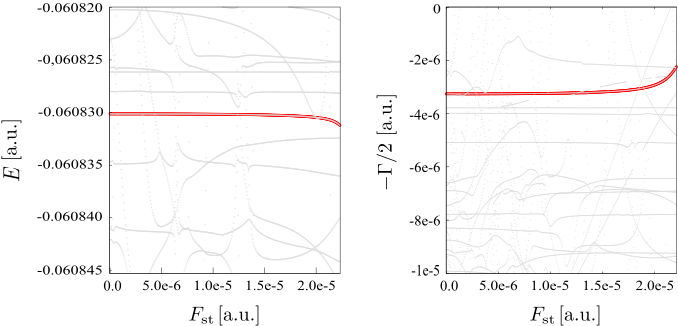}
\end{center}
\caption{Real (left) and imaginary (right) parts of the quasienergy of the Floquet state ${\cal W}^6_2$ as functions of the static field strength, for fixed periodic driving: $F = 1.0 \times 10^{-6}\,\rm a.u.$, $\omega = 0.00089\,\rm a.u.$}
\label{fig:evolEnergyvsStaticField}
\end{figure}

Figure~\ref{fig:evolEnergyvsStaticField} shows the variation of the real and imaginary parts of the quasienergy of ${\cal W}^6_2$ as a function of the static field.
The decay rate remains nearly constant up to $F_{\textrm{st}} \approx 1.5 \times 10^{-5}\,\rm a.u.$, then decreases as the field approaches the ionization limit.
The maximum value of the half decay rate is $\Gamma/2 \simeq 3.26 \times 10^{-6}\,\rm a.u.$, occurring at $F_{\textrm{st}} = 0$.

\begin{figure}
\begin{center}
\includegraphics[width=0.48\textwidth]{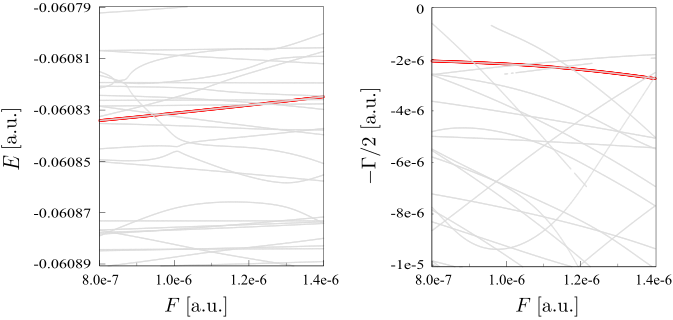}
\end{center}
\caption{Real (left) and imaginary (right) parts of the quasienergy of the Floquet state ${\cal W}^6_2$ as functions of the periodic field amplitude, at fixed static field $F_{\textrm{st}} = 2.23 \times 10^{-5}\,\rm a.u.$}
\label{fig:evolEnergyvsPeriodicField}
\end{figure}

Complementarily, Figure~\ref{fig:evolEnergyvsPeriodicField} displays the dependence of the quasienergy on the periodic field amplitude at fixed $F_{\textrm{st}} = 2.23 \times 10^{-5}\,\rm a.u.$
Here, the nondispersive character of the wave packet emerges starting from $F \approx 0.8 \times 10^{-7}\,\rm a.u.$, slightly below the lower bound observed in the purely driven case.
The state ${\cal W}^6_2$ can be tracked up to $F \approx 1.4 \times 10^{-6}\,\rm a.u.$, beyond which its identification becomes ambiguous due to several Floquet states exhibit comparable overlaps with 
the frozen planet state and a dense network of avoided crossings in the quasienergy spectrum.
Within this interval, the decay rate decreases slowly, starting from $\Gamma/2 \simeq 2.0 \times 10^{-6}\,\rm a.u.$ at the lowest periodic amplitude.

Complementarily, Figure~\ref{fig:evolEnergyvsPeriodicField} shows the {dependence} of the quasienergy on the periodic field amplitude {at fixed static field strength} $F_{\textrm{st}} = 2.23 \times 10^{-5}\,\mathrm{a.u.}$
In this case, the nondispersive character of the wave packet emerges at field amplitudes around $F \approx 0.8 \times 10^{-7}\,\mathrm{a.u.}$, slightly below the lower bound observed in the purely driven scenario.
The state ${\cal W}^6_2$ can be followed up to $F \approx 1.4 \times 10^{-6}\,\mathrm{a.u.}$, beyond which its identification becomes ambiguous due to multiple Floquet states exhibiting comparable overlaps with the frozen planet state, and a dense sequence of avoided crossings in the quasienergy spectrum.
Within this range, the decay rate decreases gradually, starting from $\Gamma/2 \simeq 2.0 \times 10^{-6}\,\mathrm{a.u.}$ at the lowest periodic field amplitude.

These results define a bounded region in parameter space $(F, F_{\textrm{st}})$ where ${\cal W}^6_2$ remains stable, localized, and long-lived.
Beyond this domain, either the decay rate increases or the wave packet delocalizes due to mixing with neighboring resonances.
While numerical diagonalization can be performed systematically, identifying the correct Floquet state requires visual inspection of Husimi projections, limiting the resolution of stability charts.

A more systematic mapping could be achieved by using predictor-corrector schemes to adiabatically follow the state across small steps in parameter space, or machine-learning tools to classify Husimi distributions.
Although such approaches remain to be implemented, they offer a promising route to improve the automation and robustness of NDWP identification.

Whereas a complete parameter scan remains unfeasible, our analysis suggests that adding a static field component can extend the lifetime of the nondispersive wave packet by up to 60\%—from approximately 22 
to 35 field cycles—based on the minimum decay rate observed in Figure~\ref{fig:evolEnergyvsStaticField} and the maximum decay rate observed in Figure~\ref{fig:evolEnergyvsPeriodicField}, 
while preserving its phase space localization.
Stability is ensured within the range $F_{\textrm{st}} \lesssim 2.3 \times 10^{-5}\,\rm a.u.$\ and $0.8 \times 10^{-7} \lesssim F \lesssim 1.4 \times 10^{-6}\,\rm a.u.$
While the present analysis focused on the $N=6$ ionization threshold, the same principles apply to other series and configurations exhibiting near-resonant coupling between 
frozen planet states. This methodology, based on monitoring the static-field response of dominant FPS components and their phase space localization, provides a practical route to engineer robust two-electron 
wave packets in more complex systems.

\subsection{Prospects for experimental detection of nondispersive wave packets}
\label{subsec:ndwp-experiment}

To date, no conclusive experimental evidence has been reported for the existence of frozen planet states in helium.
While some signatures consistent with FPS have been reported in the context of attosecond ionization 
experiments \cite{eichmann1,heber1,camus1}, their unambiguous identification remains elusive. 
Previous theoretical investigations on planar helium \cite{guzman15,gonzalez15} have shown that FPS do not leave strong signatures in the photoionization cross section from the ground state. 
However, they exhibit significant coupling to certain excited bound states, suggesting that indirect population through intermediate resonances may be feasible.

\begin{figure}
\begin{center}
\includegraphics[width=0.48\textwidth]{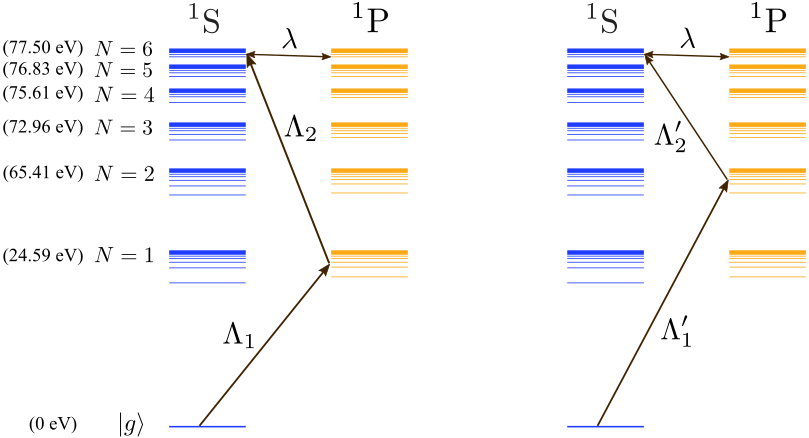}
\end{center}
\caption{Two-photon excitation pathways from the ground state to the frozen planet state $\isotope[1]{S}{\cal F}^{\,6}_2$ with energy $E= -0.06084080\,\rm a.u.$ (77.355 eV), mediated by a singly 
excited state (left) and a doubly excited state (right). The required wavelengths are:
$50.4,\textrm{nm}\leq \Lambda_1\leq 58.4,\textrm{nm}$, $22.1,\textrm{nm}\leq \Lambda_2\leq 23.5,\textrm{nm}$, $19.0,\textrm{nm}\leq \Lambda'_1\leq 19.3,\textrm{nm}$, 
and $93.7,\textrm{nm}\leq \Lambda'_2\leq 103.8,\textrm{nm}$. The final step to the NDWP ${\cal W}^{6}_2$ is achieved via near-resonant coupling at $\lambda=51776.5,\textrm{nm}$.}
\label{fig:Acoples}
\end{figure}

These findings point to a possible preparation scheme based on two-photon transitions mediated by excited states. 
Figure~\ref{fig:Acoples} illustrates two such schemes to populate the frozen planet state $\isotope[1]{S}{\cal F}^{,6}_2$ starting from the ground state: 
one involving an intermediate singly excited state, and the other a doubly excited state. 
In both cases, a subsequent near-resonant coupling to $\isotope[1]{P}{\cal F}^{,6}_1$ could lead to the formation of a nondispersive wave packet.

The required transition wavelengths lie in the far-ultraviolet (FUV) region. Although laser sources in this spectral range are available, 
the spectroscopic resolution needed to selectively address individual FPS remains beyond current experimental capabilities.

\begin{figure}
\begin{center}
\includegraphics[width=0.48\textwidth]{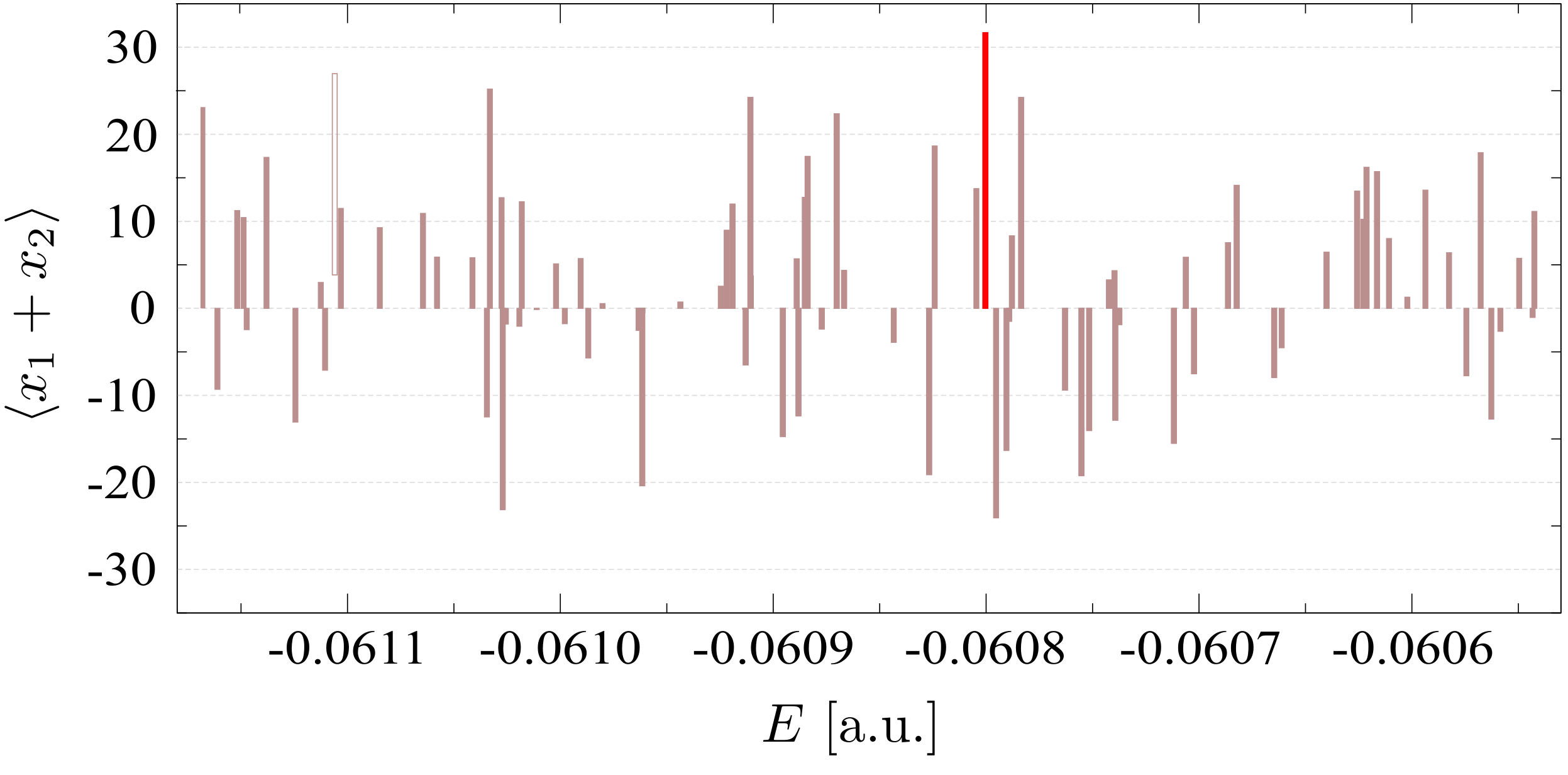}
\end{center}
\caption{Expectation value of the electronic dipole $\langle x_1+x_2\rangle$ for Floquet states at $F=3.9\times 10^{-6}\,\rm a.u.$ (intensity $1.4\times 10^{11}$ W/cm$^2$) and $\omega=0.00088\,\rm a.u.$ ($\lambda=51776.5$ nm). 
The red line indicates the dipole value for the wave packet ${\cal W}^{6}_2$ with energy $E = -0.060800\,\rm a.u.$ (77.357 eV).}
\label{fig:Dipolo}
\end{figure}

Another potential observable for the identification of NDWP is the electronic dipole moment. As shown in Figure~\ref{fig:Energias}, the Floquet state ${\cal W}^{6}_2$ displays a pronounced Stark shift 
with a positive slope as the driving amplitude increases. This behavior, also observed in planar helium \cite{GonzalezMelan2020}, reflects a large spatial polarization of the electronic density. 
Figure~\ref{fig:Dipolo} shows the expectation value of the dipole operator $\langle x_1 + x_2 \rangle$ for the Floquet eigenstates at fixed field parameters. 
The state ${\cal W}^{6}_2$ (highlighted in red) exhibits the largest dipole moment in the energy window considered, indicating that this quantity could serve as a viable experimental signature for its detection.

In summary, although the preparation and observation of NDWP in helium remain is experimentally challenging, the combination of structured excitation protocols and polarization-sensitive 
observables offers promising routes toward their realization in future ultrafast and high-resolution experiments.

\section{Summary and conclusions}
\label{sec:conclusions}

In this work, we have investigated the dynamics of frozen planet states in the full three-dimensional helium atom under periodic driving, focusing on the energy region below the sixth ionization threshold.
The analysis was performed using an efficient numerical approach that combines a spectral configuration interaction method based on Coulomb-Sturmian functions with Floquet theory and complex scaling.

Our results provide clear numerical evidence for the existence of nondispersive wave packets in three-dimensional helium, formed via near-resonant coupling between two FPS.
These wave packets remain localized along classical trajectories and exhibit relatively long lifetimes within a finite range of driving field amplitudes.
Their identification was supported by the analysis of Floquet quasienergy spectra, phase space Husimi distributions, and overlaps with field-free atomic states.

We further analyzed the effect of an additional static electric field on the stability of FPS and NDWP, and demonstrated that, for suitable combinations of static and periodic fields, the NDWP lifetime can be enhanced.
This stabilization originates from the field-induced suppression of the decay rate of the dominant FPS component of the wave packet.
However, it is confined to a restricted region of parameter space, beyond which the NDWP delocalizes and mixes with nearby resonances.
Although our analysis focused on the sixth ionization threshold, similar stabilization mechanisms are expected to arise near other thresholds.

Finally, we discussed possible strategies for the experimental detection of NDWP, including multistep excitation schemes and dipole-sensitive observables.
In particular, the strong spatial polarization of these states makes them promising targets for high-resolution attosecond spectroscopy.

Taken together, these findings represent a step forward in the theoretical understanding of strong electronic correlation under periodic driving and open the possibility for the controlled generation of long-lived, coherent two-electron wave packets in real atoms.

\acknowledgments

A.G. gratefully acknowledges support from COLCIENCIAS (Scholarship Program No. 
6172) and contract No. 71003; J.M. thanks support from the Colombian Science, Technology and Innovation 
Fund-General Royalties System (Fondo CTeI-SGR) contract No. BPIN 2013000100007; We acknowledge support from the Universidad del Valle (Proyecto de convocatoria interna CI 71102).

\bibliographystyle{apsrev} 
\bibliography{references}

\begin{thebibliography}{57}
\expandafter\ifx\csname natexlab\endcsname\relax\def\natexlab#1{#1}\fi
\expandafter\ifx\csname bibnamefont\endcsname\relax
  \def\bibnamefont#1{#1}\fi
\expandafter\ifx\csname bibfnamefont\endcsname\relax
  \def\bibfnamefont#1{#1}\fi
\expandafter\ifx\csname citenamefont\endcsname\relax
  \def\citenamefont#1{#1}\fi
\expandafter\ifx\csname url\endcsname\relax
  \def\url#1{\texttt{#1}}\fi
\expandafter\ifx\csname urlprefix\endcsname\relax\def\urlprefix{URL }\fi
\providecommand{\bibinfo}[2]{#2}
\providecommand{\eprint}[2][]{\url{#2}}

\bibitem[{\citenamefont{Ossiander et~al.}(2017)\citenamefont{Ossiander,
  Siegrist, Shirvanyan, Pazourek, Sommer, Latka, Guggenmos, Nagele, Feist,
  Burgd\"orfer et~al.}}]{Ossiander2017}
\bibinfo{author}{\bibfnamefont{M.}~\bibnamefont{Ossiander}},
  \bibinfo{author}{\bibfnamefont{F.}~\bibnamefont{Siegrist}},
  \bibinfo{author}{\bibfnamefont{V.}~\bibnamefont{Shirvanyan}},
  \bibinfo{author}{\bibfnamefont{R.}~\bibnamefont{Pazourek}},
  \bibinfo{author}{\bibfnamefont{A.}~\bibnamefont{Sommer}},
  \bibinfo{author}{\bibfnamefont{T.}~\bibnamefont{Latka}},
  \bibinfo{author}{\bibfnamefont{A.}~\bibnamefont{Guggenmos}},
  \bibinfo{author}{\bibfnamefont{S.}~\bibnamefont{Nagele}},
  \bibinfo{author}{\bibfnamefont{J.}~\bibnamefont{Feist}},
  \bibinfo{author}{\bibfnamefont{J.}~\bibnamefont{Burgd\"orfer}},
  \bibnamefont{et~al.}, \bibinfo{journal}{Nature Phys.}
  \textbf{\bibinfo{volume}{13}}, \bibinfo{pages}{280} (\bibinfo{year}{2017}).

\bibitem[{\citenamefont{Busto et~al.}(2018)\citenamefont{Busto, Barreau,
  Isinger, Turconi, Alexandridi, Harth, Zhong, Squibb, Kroon, Plogmaker
  et~al.}}]{Busto2018}
\bibinfo{author}{\bibfnamefont{D.}~\bibnamefont{Busto}},
  \bibinfo{author}{\bibfnamefont{L.}~\bibnamefont{Barreau}},
  \bibinfo{author}{\bibfnamefont{M.}~\bibnamefont{Isinger}},
  \bibinfo{author}{\bibfnamefont{M.}~\bibnamefont{Turconi}},
  \bibinfo{author}{\bibfnamefont{C.}~\bibnamefont{Alexandridi}},
  \bibinfo{author}{\bibfnamefont{A.}~\bibnamefont{Harth}},
  \bibinfo{author}{\bibfnamefont{S.}~\bibnamefont{Zhong}},
  \bibinfo{author}{\bibfnamefont{R.~J.} \bibnamefont{Squibb}},
  \bibinfo{author}{\bibfnamefont{D.}~\bibnamefont{Kroon}},
  \bibinfo{author}{\bibfnamefont{S.}~\bibnamefont{Plogmaker}},
  \bibnamefont{et~al.}, \bibinfo{journal}{J. Phys. B}
  \textbf{\bibinfo{volume}{51}}, \bibinfo{pages}{044002}
  (\bibinfo{year}{2018}).

\bibitem[{\citenamefont{Beaulieu et~al.}(2017)\citenamefont{Beaulieu, Comby,
  Clergerie, Caillat, Descamps, Dudovich, Fabre, G\'eneaux, L\'egar\'e, Petit
  et~al.}}]{Beaulieu2017}
\bibinfo{author}{\bibfnamefont{S.}~\bibnamefont{Beaulieu}},
  \bibinfo{author}{\bibfnamefont{A.}~\bibnamefont{Comby}},
  \bibinfo{author}{\bibfnamefont{A.}~\bibnamefont{Clergerie}},
  \bibinfo{author}{\bibfnamefont{J.}~\bibnamefont{Caillat}},
  \bibinfo{author}{\bibfnamefont{D.}~\bibnamefont{Descamps}},
  \bibinfo{author}{\bibfnamefont{N.}~\bibnamefont{Dudovich}},
  \bibinfo{author}{\bibfnamefont{B.}~\bibnamefont{Fabre}},
  \bibinfo{author}{\bibfnamefont{R.}~\bibnamefont{G\'eneaux}},
  \bibinfo{author}{\bibfnamefont{F.}~\bibnamefont{L\'egar\'e}},
  \bibinfo{author}{\bibfnamefont{S.}~\bibnamefont{Petit}},
  \bibnamefont{et~al.}, \bibinfo{journal}{Science}
  \textbf{\bibinfo{volume}{358}}, \bibinfo{pages}{1288} (\bibinfo{year}{2017}).

\bibitem[{\citenamefont{Yang et~al.}(2021)\citenamefont{Yang, Mainz, Rossi,
  Scheiba, Silva-Toledo, Keathley, Cirmi, and Kärtner}}]{Yang2021}
\bibinfo{author}{\bibfnamefont{Y.}~\bibnamefont{Yang}},
  \bibinfo{author}{\bibfnamefont{R.~E.} \bibnamefont{Mainz}},
  \bibinfo{author}{\bibfnamefont{G.~M.} \bibnamefont{Rossi}},
  \bibinfo{author}{\bibfnamefont{F.}~\bibnamefont{Scheiba}},
  \bibinfo{author}{\bibfnamefont{M.~A.} \bibnamefont{Silva-Toledo}},
  \bibinfo{author}{\bibfnamefont{P.~D.} \bibnamefont{Keathley}},
  \bibinfo{author}{\bibfnamefont{G.}~\bibnamefont{Cirmi}}, \bibnamefont{and}
  \bibinfo{author}{\bibfnamefont{F.~X.} \bibnamefont{Kärtner}},
  \bibinfo{journal}{Nature Communications} \textbf{\bibinfo{volume}{12}},
  \bibinfo{pages}{6641} (\bibinfo{year}{2021}).

\bibitem[{\citenamefont{Borrego-Varillas
  et~al.}(2022)\citenamefont{Borrego-Varillas, Lucchini, and
  Nisoli}}]{BorregoVarillas2022}
\bibinfo{author}{\bibfnamefont{R.}~\bibnamefont{Borrego-Varillas}},
  \bibinfo{author}{\bibfnamefont{M.}~\bibnamefont{Lucchini}}, \bibnamefont{and}
  \bibinfo{author}{\bibfnamefont{M.}~\bibnamefont{Nisoli}},
  \bibinfo{journal}{Reports on Progress in Physics}
  \textbf{\bibinfo{volume}{85}}, \bibinfo{pages}{066401}
  (\bibinfo{year}{2022}).

\bibitem[{\citenamefont{Nandi et~al.}(2022)\citenamefont{Nandi, Olofsson,
  Bertolino, Carlström, Zapata, Busto, Callegari, Fraia, Eng-Johnsson, Feifel
  et~al.}}]{Nandi2022}
\bibinfo{author}{\bibfnamefont{S.}~\bibnamefont{Nandi}},
  \bibinfo{author}{\bibfnamefont{E.}~\bibnamefont{Olofsson}},
  \bibinfo{author}{\bibfnamefont{M.}~\bibnamefont{Bertolino}},
  \bibinfo{author}{\bibfnamefont{S.}~\bibnamefont{Carlström}},
  \bibinfo{author}{\bibfnamefont{F.}~\bibnamefont{Zapata}},
  \bibinfo{author}{\bibfnamefont{D.}~\bibnamefont{Busto}},
  \bibinfo{author}{\bibfnamefont{C.}~\bibnamefont{Callegari}},
  \bibinfo{author}{\bibfnamefont{M.~D.} \bibnamefont{Fraia}},
  \bibinfo{author}{\bibfnamefont{P.}~\bibnamefont{Eng-Johnsson}},
  \bibinfo{author}{\bibfnamefont{R.}~\bibnamefont{Feifel}},
  \bibnamefont{et~al.}, \bibinfo{journal}{Nature}
  \textbf{\bibinfo{volume}{608}}, \bibinfo{pages}{488} (\bibinfo{year}{2022}).

\bibitem[{\citenamefont{Richter and Wintgen}(1993)}]{Richter1993}
\bibinfo{author}{\bibfnamefont{K.}~\bibnamefont{Richter}} \bibnamefont{and}
  \bibinfo{author}{\bibfnamefont{D.}~\bibnamefont{Wintgen}},
  \bibinfo{journal}{J. Phys. B} \textbf{\bibinfo{volume}{26}},
  \bibinfo{pages}{3719} (\bibinfo{year}{1993}).

\bibitem[{\citenamefont{Tanner et~al.}(2000)\citenamefont{Tanner, Richter, and
  M.Rost}}]{Tanner2000}
\bibinfo{author}{\bibfnamefont{G.}~\bibnamefont{Tanner}},
  \bibinfo{author}{\bibfnamefont{K.}~\bibnamefont{Richter}}, \bibnamefont{and}
  \bibinfo{author}{\bibfnamefont{J.}~\bibnamefont{M.Rost}},
  \bibinfo{journal}{Rev. Mod. Phys.} \textbf{\bibinfo{volume}{72}},
  \bibinfo{pages}{497} (\bibinfo{year}{2000}).

\bibitem[{\citenamefont{Madden and Codling}(1963)}]{Madden1963}
\bibinfo{author}{\bibfnamefont{R.~P.} \bibnamefont{Madden}} \bibnamefont{and}
  \bibinfo{author}{\bibfnamefont{K.}~\bibnamefont{Codling}},
  \bibinfo{journal}{Phys. Rev. Lett.} \textbf{\bibinfo{volume}{10}},
  \bibinfo{pages}{516} (\bibinfo{year}{1963}).

\bibitem[{\citenamefont{Byun et~al.}(2007)\citenamefont{Byun, Choi, Lee, and
  Tanner}}]{Byun2007}
\bibinfo{author}{\bibfnamefont{C.~W.} \bibnamefont{Byun}},
  \bibinfo{author}{\bibfnamefont{N.~N.} \bibnamefont{Choi}},
  \bibinfo{author}{\bibfnamefont{M.~H.} \bibnamefont{Lee}}, \bibnamefont{and}
  \bibinfo{author}{\bibfnamefont{G.}~\bibnamefont{Tanner}},
  \bibinfo{journal}{Phys. Rev. Lett.} \textbf{\bibinfo{volume}{98}},
  \bibinfo{pages}{113001} (\bibinfo{year}{2007}).

\bibitem[{\citenamefont{Ericson}(1960)}]{Ericson1960}
\bibinfo{author}{\bibfnamefont{T.}~\bibnamefont{Ericson}},
  \bibinfo{journal}{Phys. Rev. Lett.} \textbf{\bibinfo{volume}{5}},
  \bibinfo{pages}{430} (\bibinfo{year}{1960}).

\bibitem[{\citenamefont{Ericson}(1963)}]{Ericson1963}
\bibinfo{author}{\bibfnamefont{T.}~\bibnamefont{Ericson}},
  \bibinfo{journal}{Ann. Phys.} \textbf{\bibinfo{volume}{23}},
  \bibinfo{pages}{390} (\bibinfo{year}{1963}).

\bibitem[{\citenamefont{Fang et~al.}(2025)\citenamefont{Fang, Jiang, Geng,
  Březinová, Sommerlad, Tsertsvadze, Kircher, Kruse, Jahnke, Walsh
  et~al.}}]{Fang2025}
\bibinfo{author}{\bibfnamefont{Y.}~\bibnamefont{Fang}},
  \bibinfo{author}{\bibfnamefont{W.}~\bibnamefont{Jiang}},
  \bibinfo{author}{\bibfnamefont{L.}~\bibnamefont{Geng}},
  \bibinfo{author}{\bibfnamefont{I.}~\bibnamefont{Březinová}},
  \bibinfo{author}{\bibfnamefont{.}~\bibnamefont{Sommerlad}},
  \bibinfo{author}{\bibfnamefont{A.}~\bibnamefont{Tsertsvadze}},
  \bibinfo{author}{\bibfnamefont{M.}~\bibnamefont{Kircher}},
  \bibinfo{author}{\bibfnamefont{J.}~\bibnamefont{Kruse}},
  \bibinfo{author}{\bibfnamefont{T.}~\bibnamefont{Jahnke}},
  \bibinfo{author}{\bibfnamefont{N.}~\bibnamefont{Walsh}},
  \bibnamefont{et~al.}, \bibinfo{journal}{Phys. Rev. A}
  \textbf{\bibinfo{volume}{111}}, \bibinfo{pages}{023120}
  (\bibinfo{year}{2025}).

\bibitem[{\citenamefont{Blatt and Wineland}(2008)}]{Blatt2008}
\bibinfo{author}{\bibfnamefont{R.}~\bibnamefont{Blatt}} \bibnamefont{and}
  \bibinfo{author}{\bibfnamefont{D.}~\bibnamefont{Wineland}},
  \bibinfo{journal}{Nature} \textbf{\bibinfo{volume}{453}},
  \bibinfo{pages}{1008} (\bibinfo{year}{2008}).

\bibitem[{\citenamefont{Georgescu et~al.}(2014)\citenamefont{Georgescu, Ashhab,
  and Nori}}]{Georgescu2014}
\bibinfo{author}{\bibfnamefont{I.~M.} \bibnamefont{Georgescu}},
  \bibinfo{author}{\bibfnamefont{S.}~\bibnamefont{Ashhab}}, \bibnamefont{and}
  \bibinfo{author}{\bibfnamefont{F.}~\bibnamefont{Nori}},
  \bibinfo{journal}{Rev. Mod. Phys.} \textbf{\bibinfo{volume}{86}},
  \bibinfo{pages}{153} (\bibinfo{year}{2014}).

\bibitem[{\citenamefont{Ma et~al.}(2020)\citenamefont{Ma, Govoni, and
  Galli}}]{Ma2020}
\bibinfo{author}{\bibfnamefont{H.}~\bibnamefont{Ma}},
  \bibinfo{author}{\bibfnamefont{M.}~\bibnamefont{Govoni}}, \bibnamefont{and}
  \bibinfo{author}{\bibfnamefont{G.}~\bibnamefont{Galli}},
  \bibinfo{journal}{npj Comput. Mater.} \textbf{\bibinfo{volume}{6}},
  \bibinfo{pages}{85} (\bibinfo{year}{2020}).

\bibitem[{\citenamefont{Richter and Wintgen}(1990)}]{Richter1990}
\bibinfo{author}{\bibfnamefont{K.}~\bibnamefont{Richter}} \bibnamefont{and}
  \bibinfo{author}{\bibfnamefont{D.}~\bibnamefont{Wintgen}},
  \bibinfo{journal}{Phys. Rev. Lett.} \textbf{\bibinfo{volume}{65}},
  \bibinfo{pages}{1965} (\bibinfo{year}{1990}).

\bibitem[{\citenamefont{Wintgen et~al.}(1992)\citenamefont{Wintgen, Richter,
  and Tanner}}]{Richter1992}
\bibinfo{author}{\bibfnamefont{D.}~\bibnamefont{Wintgen}},
  \bibinfo{author}{\bibfnamefont{K.}~\bibnamefont{Richter}}, \bibnamefont{and}
  \bibinfo{author}{\bibfnamefont{G.}~\bibnamefont{Tanner}},
  \bibinfo{journal}{Chaos} \textbf{\bibinfo{volume}{2}}, \bibinfo{pages}{19}
  (\bibinfo{year}{1992}).

\bibitem[{\citenamefont{Schlagheck and Buchleitner}(1999)}]{Schlagheck1999}
\bibinfo{author}{\bibfnamefont{P.}~\bibnamefont{Schlagheck}} \bibnamefont{and}
  \bibinfo{author}{\bibfnamefont{A.}~\bibnamefont{Buchleitner}},
  \bibinfo{journal}{Europhys. Lett.} \textbf{\bibinfo{volume}{46}},
  \bibinfo{pages}{24} (\bibinfo{year}{1999}).

\bibitem[{\citenamefont{Buchleitner et~al.}(2002)\citenamefont{Buchleitner,
  Delande, and Zakrzewski}}]{Buchleitner2002}
\bibinfo{author}{\bibfnamefont{A.}~\bibnamefont{Buchleitner}},
  \bibinfo{author}{\bibfnamefont{D.}~\bibnamefont{Delande}}, \bibnamefont{and}
  \bibinfo{author}{\bibfnamefont{J.}~\bibnamefont{Zakrzewski}},
  \bibinfo{journal}{Phys. Rep.} \textbf{\bibinfo{volume}{368}},
  \bibinfo{pages}{409} (\bibinfo{year}{2002}).

\bibitem[{\citenamefont{Madro{\~n}ero and Buchleitner}(2008)}]{Madronero2008}
\bibinfo{author}{\bibfnamefont{J.}~\bibnamefont{Madro{\~n}ero}}
  \bibnamefont{and}
  \bibinfo{author}{\bibfnamefont{A.}~\bibnamefont{Buchleitner}},
  \bibinfo{journal}{Phys. Rev. A} \textbf{\bibinfo{volume}{77}},
  \bibinfo{pages}{053402} (\bibinfo{year}{2008}).

\bibitem[{\citenamefont{Gonzalez-Melan and
  Madro{\~n}ero}(2020)}]{GonzalezMelan2020}
\bibinfo{author}{\bibfnamefont{A.}~\bibnamefont{Gonzalez-Melan}}
  \bibnamefont{and}
  \bibinfo{author}{\bibfnamefont{J.}~\bibnamefont{Madro{\~n}ero}},
  \bibinfo{journal}{Phys. Rev. A} \textbf{\bibinfo{volume}{101}},
  \bibinfo{pages}{013414} (\bibinfo{year}{2020}).

\bibitem[{\citenamefont{Rupprecht et~al.}(2024)\citenamefont{Rupprecht, Puskar,
  N., and Leone}}]{Rupprecht2024}
\bibinfo{author}{\bibfnamefont{P.}~\bibnamefont{Rupprecht}},
  \bibinfo{author}{\bibfnamefont{N.~G.} \bibnamefont{Puskar}},
  \bibinfo{author}{\bibfnamefont{D.~M.} \bibnamefont{N.}}, \bibnamefont{and}
  \bibinfo{author}{\bibfnamefont{S.~R.} \bibnamefont{Leone}},
  \bibinfo{journal}{Phys. Rev. Research} \textbf{\bibinfo{volume}{6}},
  \bibinfo{pages}{043100} (\bibinfo{year}{2024}).

\bibitem[{\citenamefont{Floquet}(1883)}]{Floquet1}
\bibinfo{author}{\bibfnamefont{M.~G.} \bibnamefont{Floquet}},
  \bibinfo{journal}{Annales de l'\'Ecole Normale Sup\'erieure}
  \textbf{\bibinfo{volume}{12}}, \bibinfo{pages}{47} (\bibinfo{year}{1883}).

\bibitem[{\citenamefont{Shirley}(1965)}]{shirley1}
\bibinfo{author}{\bibfnamefont{J.~H.} \bibnamefont{Shirley}},
  \bibinfo{journal}{Phys. Rev.} \textbf{\bibinfo{volume}{138}},
  \bibinfo{pages}{B979} (\bibinfo{year}{1965}).

\bibitem[{\citenamefont{Kamta}(1999)}]{lagmago:phd99}
\bibinfo{author}{\bibfnamefont{G.~L.} \bibnamefont{Kamta}}, Ph.D. thesis,
  \bibinfo{school}{Universite Nationale du Benin} (\bibinfo{year}{1999}).

\bibitem[{\citenamefont{Eiglsperger et~al.}(2009)\citenamefont{Eiglsperger,
  Piraux, and Madro$\rm\tilde{n}$ero}}]{eiglsperger:3d}
\bibinfo{author}{\bibfnamefont{J.}~\bibnamefont{Eiglsperger}},
  \bibinfo{author}{\bibfnamefont{B.}~\bibnamefont{Piraux}}, \bibnamefont{and}
  \bibinfo{author}{\bibfnamefont{J.}~\bibnamefont{Madro$\rm\tilde{n}$ero}},
  \bibinfo{journal}{Phys. Rev. A} \textbf{\bibinfo{volume}{80}},
  \bibinfo{pages}{022511} (\bibinfo{year}{2009}).

\bibitem[{\citenamefont{Eiglsperger}(2010)}]{eiglspergerdiss}
\bibinfo{author}{\bibfnamefont{J.}~\bibnamefont{Eiglsperger}},
  \bibinfo{type}{Dissertation}, \bibinfo{school}{Technische Universit\"at
  M\"unchen} (\bibinfo{year}{2010}).

\bibitem[{\citenamefont{Schlagheck and Buchleitner}(1998)}]{Schlagheck1998}
\bibinfo{author}{\bibfnamefont{P.}~\bibnamefont{Schlagheck}} \bibnamefont{and}
  \bibinfo{author}{\bibfnamefont{A.}~\bibnamefont{Buchleitner}},
  \bibinfo{journal}{J. Phys. B.} \textbf{\bibinfo{volume}{31}},
  \bibinfo{pages}{L489} (\bibinfo{year}{1998}).

\bibitem[{\citenamefont{Grozdanov et~al.}(2020)\citenamefont{Grozdanov, Gusev,
  Solov’ev, and Vinitsky}}]{Grozdanov2020}
\bibinfo{author}{\bibfnamefont{T.~P.} \bibnamefont{Grozdanov}},
  \bibinfo{author}{\bibfnamefont{A.~A.} \bibnamefont{Gusev}},
  \bibinfo{author}{\bibfnamefont{E.~A.} \bibnamefont{Solov’ev}},
  \bibnamefont{and} \bibinfo{author}{\bibfnamefont{S.~I.}
  \bibnamefont{Vinitsky}}, \bibinfo{journal}{Eur. Phys. J. D}
  \textbf{\bibinfo{volume}{74}}, \bibinfo{pages}{161} (\bibinfo{year}{2020}).

\bibitem[{\citenamefont{Foumouo et~al.}(2006)\citenamefont{Foumouo, Kamta,
  Edah, and Piraux}}]{foumouo06}
\bibinfo{author}{\bibfnamefont{E.}~\bibnamefont{Foumouo}},
  \bibinfo{author}{\bibfnamefont{G.~L.} \bibnamefont{Kamta}},
  \bibinfo{author}{\bibfnamefont{G.}~\bibnamefont{Edah}}, \bibnamefont{and}
  \bibinfo{author}{\bibfnamefont{B.}~\bibnamefont{Piraux}},
  \bibinfo{journal}{Phys. Rev. A} \textbf{\bibinfo{volume}{74}},
  \bibinfo{pages}{063409} (\bibinfo{year}{2006}).

\bibitem[{\citenamefont{Rotenberg}(1970)}]{rotenberg70}
\bibinfo{author}{\bibfnamefont{M.}~\bibnamefont{Rotenberg}},
  \bibinfo{journal}{Adv. At. Mol. Phys.} \textbf{\bibinfo{volume}{6}},
  \bibinfo{pages}{233} (\bibinfo{year}{1970}).

\bibitem[{\citenamefont{Huens et~al.}(1997)\citenamefont{Huens, Piraux,
  Bugacov, and Gajda}}]{huens97}
\bibinfo{author}{\bibfnamefont{E.}~\bibnamefont{Huens}},
  \bibinfo{author}{\bibfnamefont{B.}~\bibnamefont{Piraux}},
  \bibinfo{author}{\bibfnamefont{A.}~\bibnamefont{Bugacov}}, \bibnamefont{and}
  \bibinfo{author}{\bibfnamefont{M.}~\bibnamefont{Gajda}},
  \bibinfo{journal}{Phys. Rev. A} \textbf{\bibinfo{volume}{55}},
  \bibinfo{pages}{2132} (\bibinfo{year}{1997}).

\bibitem[{\citenamefont{Varschalovich et~al.}(2008)\citenamefont{Varschalovich,
  Moskalev, and Khersonskii}}]{Varschalovich08}
\bibinfo{author}{\bibfnamefont{D.~A.} \bibnamefont{Varschalovich}},
  \bibinfo{author}{\bibfnamefont{A.~N.} \bibnamefont{Moskalev}},
  \bibnamefont{and} \bibinfo{author}{\bibfnamefont{V.~K.}
  \bibnamefont{Khersonskii}}, \emph{\bibinfo{title}{Quantum Theory of Angular
  Momentum}} (\bibinfo{publisher}{World Scientific},
  \bibinfo{address}{Singapore}, \bibinfo{year}{2008}).

\bibitem[{\citenamefont{Delande and Zakrzewski}(1998)}]{delande8}
\bibinfo{author}{\bibfnamefont{D.}~\bibnamefont{Delande}} \bibnamefont{and}
  \bibinfo{author}{\bibfnamefont{J.}~\bibnamefont{Zakrzewski}},
  \bibinfo{journal}{Phys. Rev. A} \textbf{\bibinfo{volume}{58}},
  \bibinfo{pages}{466} (\bibinfo{year}{1998}).

\bibitem[{\citenamefont{Aguilar and Combes}(1971)}]{AC:CMP22-269}
\bibinfo{author}{\bibfnamefont{J.}~\bibnamefont{Aguilar}} \bibnamefont{and}
  \bibinfo{author}{\bibfnamefont{J.~M.} \bibnamefont{Combes}},
  \bibinfo{journal}{Comm. Math. Phys.} \textbf{\bibinfo{volume}{22}},
  \bibinfo{pages}{269} (\bibinfo{year}{1971}).

\bibitem[{\citenamefont{Balslev and Combes}(1971)}]{balslev1}
\bibinfo{author}{\bibfnamefont{E.}~\bibnamefont{Balslev}} \bibnamefont{and}
  \bibinfo{author}{\bibfnamefont{J.~M.} \bibnamefont{Combes}},
  \bibinfo{journal}{Comm. Math. Phys.} \textbf{\bibinfo{volume}{22}},
  \bibinfo{pages}{280} (\bibinfo{year}{1971}).

\bibitem[{\citenamefont{Simon}(1973)}]{S:AoM97-247}
\bibinfo{author}{\bibfnamefont{B.}~\bibnamefont{Simon}}, \bibinfo{journal}{The
  Annals of Mathematics} \textbf{\bibinfo{volume}{97}}, \bibinfo{pages}{247}
  (\bibinfo{year}{1973}).

\bibitem[{\citenamefont{Reinhardt}(1982)}]{R:ARPC33-223}
\bibinfo{author}{\bibfnamefont{W.~P.} \bibnamefont{Reinhardt}},
  \bibinfo{journal}{Annual Review of Physical Chemistry}
  \textbf{\bibinfo{volume}{33}}, \bibinfo{pages}{223} (\bibinfo{year}{1982}).

\bibitem[{\citenamefont{Ho}(1983)}]{H:PRep99-1}
\bibinfo{author}{\bibfnamefont{Y.}~\bibnamefont{Ho}}, \bibinfo{journal}{Physics
  Reports} \textbf{\bibinfo{volume}{99}}, \bibinfo{pages}{1}
  (\bibinfo{year}{1983}).

\bibitem[{\citenamefont{Lanczos}(1950)}]{lanczos1}
\bibinfo{author}{\bibfnamefont{C.}~\bibnamefont{Lanczos}}, \bibinfo{journal}{J.
  Res. Natl. Bur. Stand.} \textbf{\bibinfo{volume}{45}}, \bibinfo{pages}{225}
  (\bibinfo{year}{1950}).

\bibitem[{\citenamefont{Krug}(2001)}]{krug1}
\bibinfo{author}{\bibfnamefont{A.}~\bibnamefont{Krug}},
  \bibinfo{type}{Dissertation},
  \bibinfo{school}{Ludwig-Maximilians-Universit\"at M\"unchen}
  (\bibinfo{year}{2001}).

\bibitem[{\citenamefont{Schlagheck}(1999)}]{schlagheck1}
\bibinfo{author}{\bibfnamefont{P.}~\bibnamefont{Schlagheck}}, Ph.D. thesis,
  \bibinfo{school}{Technische Universit\"at M\"unchen} (\bibinfo{year}{1999}).

\bibitem[{\citenamefont{Lichtenberg and Lieberman}(1983)}]{lichtenberg83}
\bibinfo{author}{\bibfnamefont{A.~J.} \bibnamefont{Lichtenberg}}
  \bibnamefont{and} \bibinfo{author}{\bibfnamefont{M.~A.}
  \bibnamefont{Lieberman}}, \emph{\bibinfo{title}{Regular and Stochastic
  Motion}} (\bibinfo{publisher}{Springer-Verlag}, \bibinfo{address}{New York},
  \bibinfo{year}{1983}).

\bibitem[{\citenamefont{Ostrovsky and Prudov}(1995)}]{ostrovsky1}
\bibinfo{author}{\bibfnamefont{V.~N.} \bibnamefont{Ostrovsky}}
  \bibnamefont{and} \bibinfo{author}{\bibfnamefont{N.~V.}
  \bibnamefont{Prudov}}, \bibinfo{journal}{J. Phys. B}
  \textbf{\bibinfo{volume}{28}}, \bibinfo{pages}{4435} (\bibinfo{year}{1995}).

\bibitem[{\citenamefont{Schlagheck and Buchleitner}(2003)}]{schlagheck2}
\bibinfo{author}{\bibfnamefont{P.}~\bibnamefont{Schlagheck}} \bibnamefont{and}
  \bibinfo{author}{\bibfnamefont{A.}~\bibnamefont{Buchleitner}},
  \bibinfo{journal}{Eur. Phys. J. D} \textbf{\bibinfo{volume}{22}},
  \bibinfo{pages}{401} (\bibinfo{year}{2003}).

\bibitem[{\citenamefont{Madro$\rm\tilde{n}$ero
  et~al.}(2005)\citenamefont{Madro$\rm\tilde{n}$ero, Schlagheck, Hilico,
  Gr\'emaud, Delande, and Buchleitner}}]{madronero:epl05}
\bibinfo{author}{\bibfnamefont{J.}~\bibnamefont{Madro$\rm\tilde{n}$ero}},
  \bibinfo{author}{\bibfnamefont{P.}~\bibnamefont{Schlagheck}},
  \bibinfo{author}{\bibfnamefont{L.}~\bibnamefont{Hilico}},
  \bibinfo{author}{\bibfnamefont{B.}~\bibnamefont{Gr\'emaud}},
  \bibinfo{author}{\bibfnamefont{D.}~\bibnamefont{Delande}}, \bibnamefont{and}
  \bibinfo{author}{\bibfnamefont{A.}~\bibnamefont{Buchleitner}},
  \bibinfo{journal}{Europhys. Lett.} \textbf{\bibinfo{volume}{70}},
  \bibinfo{pages}{183} (\bibinfo{year}{2005}).

\bibitem[{\citenamefont{Madro$\rm\tilde{n}$ero and
  Buchleitner}(2008)}]{madronero08:pra}
\bibinfo{author}{\bibfnamefont{J.}~\bibnamefont{Madro$\rm\tilde{n}$ero}}
  \bibnamefont{and}
  \bibinfo{author}{\bibfnamefont{A.}~\bibnamefont{Buchleitner}},
  \bibinfo{journal}{Phys. Rev. A} \textbf{\bibinfo{volume}{77}},
  \bibinfo{pages}{053402} (\bibinfo{year}{2008}).

\bibitem[{\citenamefont{Richter et~al.}(1992)\citenamefont{Richter, Briggs,
  Wintgen, and Solovev}}]{richter6}
\bibinfo{author}{\bibfnamefont{K.}~\bibnamefont{Richter}},
  \bibinfo{author}{\bibfnamefont{J.~S.} \bibnamefont{Briggs}},
  \bibinfo{author}{\bibfnamefont{D.}~\bibnamefont{Wintgen}}, \bibnamefont{and}
  \bibinfo{author}{\bibfnamefont{E.~A.} \bibnamefont{Solovev}},
  \bibinfo{journal}{J. Phys. B} \textbf{\bibinfo{volume}{25}},
  \bibinfo{pages}{3929} (\bibinfo{year}{1992}).

\bibitem[{\citenamefont{Eichmann et~al.}(1990)\citenamefont{Eichmann, Lange,
  and Sandner}}]{eichmann1}
\bibinfo{author}{\bibfnamefont{U.}~\bibnamefont{Eichmann}},
  \bibinfo{author}{\bibfnamefont{V.}~\bibnamefont{Lange}}, \bibnamefont{and}
  \bibinfo{author}{\bibfnamefont{W.}~\bibnamefont{Sandner}},
  \bibinfo{journal}{Phys. Rev. Lett.} \textbf{\bibinfo{volume}{64}},
  \bibinfo{pages}{274} (\bibinfo{year}{1990}).

\bibitem[{\citenamefont{Heber et~al.}(1997)\citenamefont{Heber, Seng, Halka,
  Eichmann, and Sandner}}]{heber1}
\bibinfo{author}{\bibfnamefont{K.~D.} \bibnamefont{Heber}},
  \bibinfo{author}{\bibfnamefont{M.}~\bibnamefont{Seng}},
  \bibinfo{author}{\bibfnamefont{M.}~\bibnamefont{Halka}},
  \bibinfo{author}{\bibfnamefont{U.}~\bibnamefont{Eichmann}}, \bibnamefont{and}
  \bibinfo{author}{\bibfnamefont{W.}~\bibnamefont{Sandner}},
  \bibinfo{journal}{Phys. Rev. A} \textbf{\bibinfo{volume}{56}},
  \bibinfo{pages}{1255} (\bibinfo{year}{1997}).

\bibitem[{\citenamefont{Camus et~al.}(1989)\citenamefont{Camus, Gallagher,
  Lecomte, Pillet, Pruvost, and Boulmer}}]{camus1}
\bibinfo{author}{\bibfnamefont{P.}~\bibnamefont{Camus}},
  \bibinfo{author}{\bibfnamefont{T.~F.} \bibnamefont{Gallagher}},
  \bibinfo{author}{\bibfnamefont{J.~M.} \bibnamefont{Lecomte}},
  \bibinfo{author}{\bibfnamefont{P.}~\bibnamefont{Pillet}},
  \bibinfo{author}{\bibfnamefont{L.}~\bibnamefont{Pruvost}}, \bibnamefont{and}
  \bibinfo{author}{\bibfnamefont{J.}~\bibnamefont{Boulmer}},
  \bibinfo{journal}{Phys. Rev. Lett.} \textbf{\bibinfo{volume}{62}},
  \bibinfo{pages}{2365} (\bibinfo{year}{1989}).

\bibitem[{\citenamefont{Leopold and Percival}(1980)}]{percival1}
\bibinfo{author}{\bibfnamefont{J.~G.} \bibnamefont{Leopold}} \bibnamefont{and}
  \bibinfo{author}{\bibfnamefont{I.~C.} \bibnamefont{Percival}},
  \bibinfo{journal}{J. Phys. B} \textbf{\bibinfo{volume}{13}},
  \bibinfo{pages}{1037} (\bibinfo{year}{1980}).

\bibitem[{\citenamefont{Burgers et~al.}(1995)\citenamefont{Burgers, Wintgen,
  and Rost}}]{burgers1995}
\bibinfo{author}{\bibfnamefont{A.}~\bibnamefont{Burgers}},
  \bibinfo{author}{\bibfnamefont{D.}~\bibnamefont{Wintgen}}, \bibnamefont{and}
  \bibinfo{author}{\bibfnamefont{J.-M.} \bibnamefont{Rost}},
  \bibinfo{journal}{J. Phys. B: At. Mol. Opt. Phys.}
  \textbf{\bibinfo{volume}{28}}, \bibinfo{pages}{3163} (\bibinfo{year}{1995}).

\bibitem[{\citenamefont{J.~M.~Rost and Kaindl}(1997)}]{rost1997}
\bibinfo{author}{\bibfnamefont{M.~D.} \bibnamefont{J.~M.~Rost},
  \bibfnamefont{K.~Schulz}} \bibnamefont{and}
  \bibinfo{author}{\bibfnamefont{G.}~\bibnamefont{Kaindl}},
  \bibinfo{journal}{J. Phys. B: At. Mol. Opt. Phys.}
  \textbf{\bibinfo{volume}{30}}, \bibinfo{pages}{4663} (\bibinfo{year}{1997}).

\bibitem[{\citenamefont{Guzm{\'a}n}(2015)}]{guzman15}
\bibinfo{author}{\bibfnamefont{I.}~\bibnamefont{Guzm{\'a}n}},
  \bibinfo{type}{Undergraduate thesis}, \bibinfo{school}{Universidad del Valle}
  (\bibinfo{year}{2015}).

\bibitem[{\citenamefont{Gonz\'alez et~al.}(2015)\citenamefont{Gonz\'alez,
  Guzm\'an, and Madro$\rm\tilde{n}$ero}}]{gonzalez15}
\bibinfo{author}{\bibfnamefont{A.}~\bibnamefont{Gonz\'alez}},
  \bibinfo{author}{\bibfnamefont{I.}~\bibnamefont{Guzm\'an}}, \bibnamefont{and}
  \bibinfo{author}{\bibfnamefont{J.}~\bibnamefont{Madro$\rm\tilde{n}$ero}},
  \bibinfo{journal}{J. Phys. Conf. Ser.} \textbf{\bibinfo{volume}{635}},
  \bibinfo{pages}{092088} (\bibinfo{year}{2015}).

\end{thebibliography}

\end{document}